\documentclass[onecolumn,11pt]{article}

\usepackage{setspace}
\usepackage{amsmath}
\usepackage{authblk}
\usepackage{graphicx}
\usepackage[utf8]{inputenc}
\usepackage{algorithm}
\usepackage{algpseudocode}
\usepackage{subcaption}
\usepackage[percent]{overpic} 
\usepackage[capitalize]{cleveref}
\usepackage{printlen}
\uselengthunit{in} 

\usepackage{xcolor}

\usepackage[backend=biber, style=chem-angew]{biblatex}
\usepackage{geometry}
\renewenvironment{abstract}
{
  \section*{Abstract}
}

\begin{document}


\title{AES-Debye: an Accurate, Efficient, and Scalable Engine for Debye Scattering Calculations}

\author[1,2,*]{Navid Panchi}
\author[2]{Sebastian Kuckuk}
\author[2]{Markus Wittmann}
\author[1]{Michael Engel}
\author[3,4,*]{Alberto Leonardi}
\affil[1]{Institute for Multiscale Simulation, Friedrich-Alexander-Universität Erlangen-Nürnberg, 91058 Erlangen, Germany}
\affil[2]{Erlangen National High Performance Computing Center (NHR@FAU), Friedrich-Alexander-Universität Erlangen-Nürnberg, 91058 Erlangen, Germany}
\affil[3]{Technische Fakultät, Friedrich-Alexander-Universität Erlangen-Nürnberg, 91058 Erlangen, Germany}
\affil[4]{UKRI-STFC Rutherford Appleton Laboratory, Diamond Light Source, Harwell Science \& Innovation Campus, OX110DE Didcot, United Kingdom}

\affil[*]{navid.panchi@fau.de, alberto.leonardi@fau.de}

\date{\today}

\maketitle


\begin{center}
    \includegraphics[width=0.5\columnwidth]{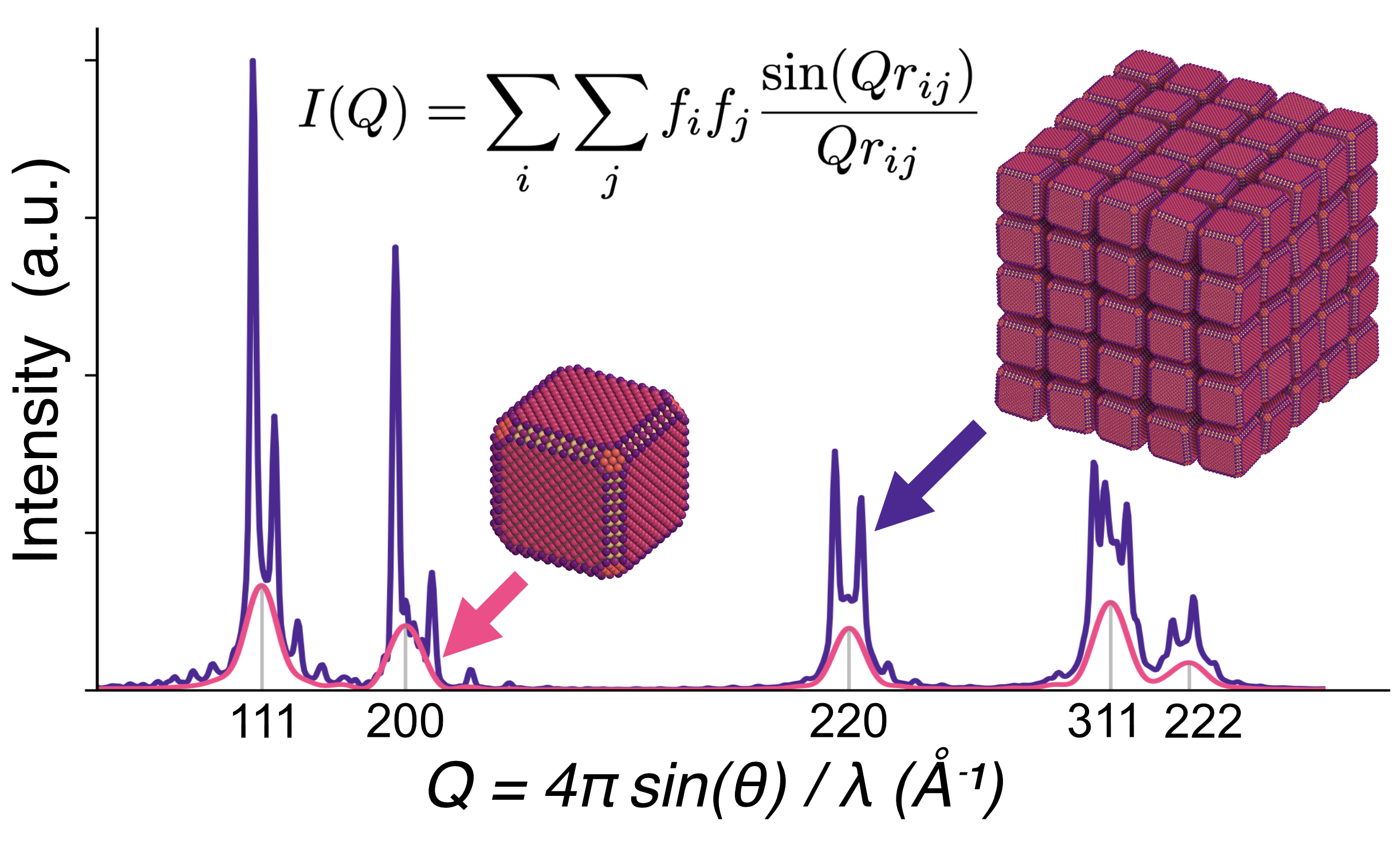}
\end{center}

\section*{Synopsis}
AES-Debye is an accurate, efficient, and scalable engine for evaluating the Debye scattering equation that enables total scattering calculations for large atomistic models. Corrected pair-distance binning and numerically robust accumulation suppress discretization artifacts, while parallel CPU/GPU execution enables efficient computation of scattering intensities and pair distribution functions for structural analysis.


\begin{abstract}
Total scattering models are essential for characterizing the structure and disorder of nanoscale materials. The Debye scattering equation (DSE) provides a rigorous route to elastic total scattering, but its direct evaluation is computationally demanding because pairwise contributions must be accumulated at every scattering vector, whereas common acceleration strategies based on binned pair-distance distributions or gridded fast Fourier transforms can introduce discretization and aliasing artifacts that compromise diffuse-scattering accuracy. Here, we present AES-Debye, an accuracy-preserving DSE framework that aggregates pair distances into a pair distribution function (PDF) using corrected bin centers and numerically robust accumulation to suppress discretization and summation errors. A data-locality-aware parallel design enables efficient execution on CPUs and GPUs. We demonstrate strong scalability by computing a high-resolution total scattering profile for a system of 90 million atoms, $(0.1,\mu\mathrm{m})^{3}$, in minutes on a distributed-memory CPU platform. These capabilities extend accurate elastic total scattering calculations to large, complex systems while simultaneously providing high-resolution PDFs for downstream structural analysis.
\end{abstract}


\section{Introduction}

Powder diffraction is widely used for characterizing the structure and microstructure of materials~\cite{Kaduk2021}.
Experimental diffraction profiles are analyzed using a range of modeling approaches to extract statistically meaningful information about the sample.
Extracting this information is particularly critical for nanostructured materials, where structural features such as crystalline domain size, shape, and lattice distortions dictate macroscopic physical and chemical properties~\cite{Zhang2011, Cernuto2011,latticeDistortionChemProp}.
Because these characteristics manifest as line broadening and diffuse scattering, robust total scattering models are essential for reliable interpretation.

Traditionally, powder diffraction data have been analyzed using line profile methods such as Rietveld refinement~\cite{rietveld1967line} and whole powder pattern modeling (WPPM)~\cite{wppm}.  
These approaches determine peak positions and relative integrated intensities from the crystal structure, while peak shapes are described by parameterized functions whose parameters are optimized against experimental data.  
Although computationally efficient, such methods rely on the Bragg approximation within long-range ordered domains and therefore do not fully capture the structural disorder responsible for diffuse scattering.  
Peak broadening from lattice distortions is typically incorporated through approximate models, that are specialized for specific type of disorder~\cite{Gelisio2016}. Moreover, only a limited number of such models are available for highly disordered materials.

To compute powder total scattering profiles while accounting for arbitrary structural disorder, two broad classes of approaches are commonly used: calculations based on the Fourier transform of the scattering density and evaluations of the Debye scattering equation (DSE). FFT-based methods typically exhibit favorable asymptotic scaling: $O(M\log M)$ with the number of grid points M, and are widely used for diffraction calculations of large systems. However, for powder scattering the orientational averaging required to obtain the one-dimensional profile requires interpolation between Cartesian reciprocal space grids and spherical shells ~\cite{Ross2014, Kieffer2013}. Moreover, FFT-based approaches require discretization of the scattering density on a finite real space grid, making the accuracy dependent on the chosen grid resolution and sampling strategy.
With memory requirements a critical sampling factor, the mapping of the scattering intensities from Cartesian space onto radial bins involves a non-unique rebinning procedure, with the final profile depending on the specific interpolation and pixel weighting strategy employed.
Crucially, eliminating directional sampling approximations during this step requires a number of unique angular projections that scales linearly with the number of atoms ~\cite{wpdfm}.
Sampling these dense reciprocal space directions explicitly forces the overall scaling back to $O(N^2)$, effectively neutralizing the efficiency gains of the underlying transform.
Consequently, the powder pattern represents an approximation whose accuracy depends on the reciprocal-space sampling and integration strategy, with discrepancies generally becoming more pronounced at high Q.
In contrast, the DSE operates directly on atomic coordinates and avoids density discretization, at the expense of a formally $O(N^2)$ pairwise summation ~\cite{Debye1915}. Without assuming periodicity or long-range order, it is well suited to disordered and nanoscale systems~\cite{scardi2016vibrational, Gelisio2016}.

In its common form, the Debye scattering equation is given as: 
\begin{equation}\label{eq:1}
    I(Q) = \sum_i \sum_j f_i f_j \frac{\sin(Q r_{ij})}{Q r_{ij}},
\end{equation}
where $r_{ij} = |\mathbf{r}_j - \mathbf{r}_i|$ is the interatomic distance and $f_i$, $f_j$ are atomic scattering factors.  
The scattering vector magnitude is $Q = \frac{4\pi}{\lambda}\sin\theta$, where $\lambda$ is the incident wavelength and $\theta$ is the scattering angle~\cite{warren1990}.
This formulation mathematically enforces an exact orientational average of the powder profile.
Consequently, while the DSE fully captures the internal structure and morphological anisotropy of individual nanoparticles, including finite size and shape effects, it does not natively describe directional correlations or texture, such as those arising in aligned systems under flow, external fields, or deformation. 
Specialized extensions have been developed to incorporate preferred orientation into total scattering formalisms~\cite{cervellino_texture}, which can be included in a future extension.

Evaluating \cref{eq:1} requires one $\mathrm{sinc}(x)=\sin(x)/x$ evaluation for every atomic pair, giving $O(N^2)$ complexity for $N$ atoms.  
Since $I(Q)$ is typically sampled on many $Q$ points, the total cost increases to $O(\ell N^2)$, where $\ell$ denotes the number of sampled $Q$ values.  
In practice, this cost is compounded by the large number of transcendental evaluations and by floating point accumulation of terms spanning a wide range of magnitudes, which can introduce non-negligible numerical error.

A variety of strategies have been explored to improve efficiency:
\begin{itemize}
    \item \emph{Algorithmic modifications} reduce computational cost by changing the underlying physics or sampling strategy. These include pair distance cutoffs~\cite{debyer}, which introduce arbitrary correlation length and shape artifacts; golden ratio orientation sampling for powder averaging~\cite{watson2013rapid}, which is most accurate at low scattering angles; and methods that exploit lattice periodicity~\cite{grover2001efficient, thomas2010new, wpdfm}, which are inherently unsuitable to disordered systems.
    \item \emph{Full Debye methods} include brute force summation, sometimes accelerated on GPUs~\cite{gelisio2010real, debyecalc}, as well as PDF based approaches in which a pair distribution function is constructed first then the scattering intensity is computed~\cite{xansons, cudebye, cadishi}.
\end{itemize}
However, many existing implementations gain speed at the expense of accuracy by relying on coarse histograms or 32-bit arithmetic, and they remain susceptible to numerical errors associated with floating point accumulation and binning approximations~\cite{hallmarnot}.

Leonardi and Bish (Rose-X)~\cite{rose_x} improved numerical accuracy through fine binning, dynamic bin center correction, and integer arithmetic to suppress floating point noise. However, their implementation exhibits poor performance in disordered systems because random access into the PDF degrades memory locality and cache efficiency.  
Building on these observations, we identify the principal bottlenecks and redesign the computation around domain decomposition to achieve more local and predictable data access.  
The resulting engine, AES-Debye, retains the numerical rigor of Rose-X while introducing a hybrid OpenMP/MPI/CUDA parallel framework for CPUs and GPUs, yielding speedups of up to 20$\times$.


\section{Implementation details}

This section describes AES-Debye, whose design is centered on three objectives: accuracy, efficiency, and scalability.
\emph{Accuracy} is achieved through a PDF based formulation with bin center correction and precision aware integer accumulation, which suppresses discretization and summation artifacts.
\emph{Efficiency} is improved through cache friendly data layouts, compact histograms, strip mining, and a cell-list–based domain decomposition that enhances localized memory locality and reduces latency.
\emph{Scalability} is provided by a hybrid OpenMP/MPI/CUDA design that exploits shared memory, device level, and distributed memory parallelism across CPUs, GPUs, and multi-node systems.

\subsection{PDF formulation with bin center correction}

We reformulate \cref{eq:1} by grouping recurring pair distances into a PDF, allowing the DSE to be evaluated more efficiently.
Expressing the intensity in terms of representative distances $\nu_k$ and their multiplicities $N_k$ gives
\begin{equation}\label{eq:pdf_dse}
    I(Q) = \sum_a \sum_b \bigg[f_a f_b \sum_k N_k \frac{\sin(Q \nu_k)}{Q \nu_k}\bigg],
\end{equation}
where $a$ and $b$ denote atomic species.

This two-step formulation decouples the atomic pair enumeration from the reciprocal space evaluation.
Constructing the PDF still requires computing all interatomic distances and therefore scales as $O(N^2)$.
Once the histogram has been built, however, evaluating \cref{eq:pdf_dse} scales only as $O(\ell\, n_{\mathrm{bins}})$, where $\ell$ is the number of sampled $Q$ points and $n_{\mathrm{bins}}$ is the number of populated histogram bins.
The PDF based approach is therefore advantageous whenever $N^2 + \ell\, n_{\mathrm{bins}} < \ell N^2$. In practice, for large systems and dense $Q$-grids,
$n_{\mathrm{bins}} \ll N^2$, so the cost of reciprocal space evaluations is reduced substantially relative to direct pairwise summation at every $Q$ point.

Distinct pair distances are accumulated into uniformly spaced histogram bins of width $\Delta$.
The choice of $\Delta$ determines the numerical resolution and introduces artifacts at $Q$-intervals proportional to $1/\Delta$.
A second source of error arises when the bin center $\nu_k$ is a poor proxy for the average distance of the pairs assigned to that bin, which can lead to unphysical negative intensities~\cite{hallmarnot}.

Leonardi and Bish~\cite{rose_x} addressed this by storing, in addition to the bin counts, the accumulated difference between the squared pair distances and the squared bin center.
After the PDF is constructed, this error in squared pair distance (ESPD) is used to correct the bin center so it more accurately represents the average distance within the bin.
Let $\rho$ be the corrected center and $\nu$ the original center, and define $\delta=\rho-\nu$ and $\psi=\rho^2-\nu^2$.
From $(\nu+\delta)^2-\nu^2=\psi$, one obtains
\begin{equation}
\delta^2 + 2\nu\,\delta - \psi = 0,
\end{equation}
which admits a closed form solution for bins containing a single (monodisperse) distance.

For bins containing multiple contributing distances, a series expansion provides an accurate estimate of the mean shift,
\begin{equation}
\langle \delta_k \rangle
= \frac{1}{N_k}\!\left(\frac{\psi_k}{2\nu_k} - \frac{\psi_k^3}{4\nu_k^2} + \frac{\psi_k^4}{8\nu_k^3} + O(\ldots)\right),
\end{equation}
where $N_k$ is the number of pairs in bin $k$ and $\psi_k$ is the accumulated ESPD.
The corrected representative distance is the $\hat{\nu}_k=\nu_k+\langle\delta_k\rangle$, which replaces $\nu_k$ in \cref{eq:pdf_dse}.
Full derivations and error propagation details are given in Ref.~\cite{rose_x}.

\subsection{Numerical precision and accumulation}

Double precision floating point (\texttt{double}) provides a wide dynamic range (up to $10^{308}$) but limited mantissa precision: integers are represented exactly only up to $2^{53}\!\approx\!9\times10^{15}$.
Beyond this threshold, intermediate values can no longer by represented exactly, which compromises the accuracy of large accumulations.
Because bin center correction depends on precise tracking of the squared error term $\psi$, we perform all bin indexing and $\psi$ accumulations in 64-bit integers (\texttt{int64}).

Atomic positions are first normalized to the simulation box, mapped to the interval $[0,1)$, scaled by $10^9$, and stored as \texttt{int64}.
This preserves the required spatial precision.
For example, coordinates typically contain no more than six significant digits in common simulation outputs from LAMMPS~\cite{lammps} and HOOMD-blue~\cite{hoomd}.
The resulting maximum squared interatomic distance is on the order of $(10^9)^2$, which remains well within the \texttt{int64} range, so $\psi$ can be accumulated exactly up to \texttt{INT64\_MAX}.
To handle extreme cases safely, we record potential over- and underflow events in an auxiliary PDF.
This introduces one additional conditional per update while preserving correctness.

For uniformly spaced PDF bins, the worst case accumulation occurs in the final bin, where the center $\nu$ is farthest from the upper boundary $\nu_{\text{ul}}$.
If all contributing pairs lie exactly at $\nu_{\text{ul}}$, the maximum safe count is
\begin{equation}
n_\text{max} = \frac{\texttt{INT64\_MAX}}{\nu_{\text{ul}} - \nu},
\end{equation}
which is typically $\approx 5.6\times 10^6$ for the bin widths used here.
Because explicit overflow checks on every update would introduce branching and misprediction overhead, we instead maintain a per-bin counter initialized to $n_\text{max}$ and decrement it with each update.
When the counter reaches zero, a single overflow check is performed on $\psi$.
If an overflow is detected, it is recorded and the counter is reset to $n_\text{max}$; otherwise the counter is refreshed according to the remaining safe headroom.
This strategy preserves numerical robustness while keeping runtime overhead low.

\subsection{Data structures and memory layout}

Atomic positions are stored in a structure-of-arrays (SoA) layout, with three separate arrays for $x$, $y$, and $z$, so that sequential coordinate access ($i, i{+}1,\dots$) remains cache friendly and does not become a bottleneck.
Profiling with Intel VTune shows that the dominant CPU limitation is latency from random updates to the PDF histogram; the same access pattern is also a major performance limiter on GPUs.
Because memory access becomes progressively slower from L1 to  L2 to L3 caches and finally to DRAM, large high-accuracy PDFs often exceed L1/L2 capacity and reside in L3 or main memory.
Structural disorder worsens this effect by spreading updates across many bins and thereby increasing the number of random accesses.
As a result, crystalline systems run fastest, whereas disordered systems tend to become latency bound.

To mitigate this behavior, we employ two complementary strategies:

\begin{itemize}
    \item \emph{Strip mining (CPU only).}
    We buffer bin indices $k$ and ESPD values $\psi$ in two small contiguous arrays of size 512: $k$ is stored in 32-bit unsigned integers (\texttt{uint32}), and $\psi$ in \texttt{int64}.
    Once full, the buffers are flushed to the global PDF in a dedicated update loop.
    This form of write combining reduces random stores and improves cache locality, consistent with earlier observations~\cite{cadishi}.

    \item \emph{Reducing memory footprint.}
    To lower access latency, each PDF bin is compacted to 128 bits: \texttt{int64} for $\psi$, together with \texttt{int32} values for the count $N_k$ and the per-bin headroom counter $n_\text{max}$.
    When $n_\text{max}$ reaches zero, an overflow or underflow check is triggered for $\psi$; if necessary, the current $N_k$ is offloaded to a secondary \texttt{int64} accumulator.
    On GPUs, pair count overflows are instead tracked with a separate auxiliary counter to avoid large scattered writes.
\end{itemize}

\subsection{Cell-list–based domain decomposition}

To improve memory access patterns during PDF construction, we partition the simulation box into a regular grid of equally sized cells.  
A prescribed number of cells is defined along each spatial dimension, atoms are assigned to cells according to their position. The coordinates are reordered by cell so that atoms belonging to the same cell occupy contiguous memory.
For spatially non-uniform systems, the number of cells along each axis can be scaled in proportion to the corresponding system dimensions, promoting a more even distribution of atom pairs across cells and thereby improving load balance.

\begin{figure}
    \centering
    \includegraphics[width=0.5\columnwidth]{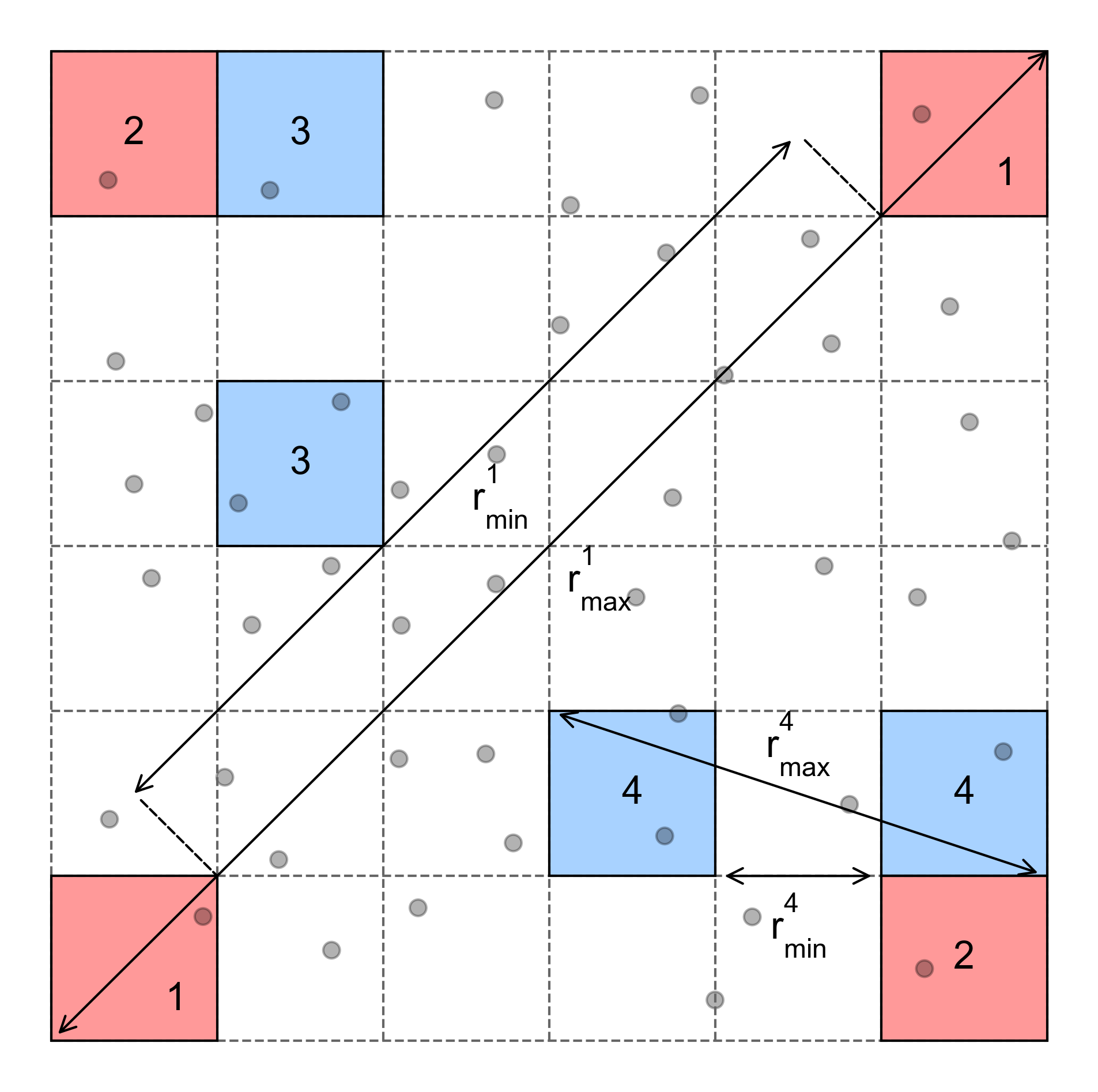}
    \includegraphics[width=0.5\columnwidth]{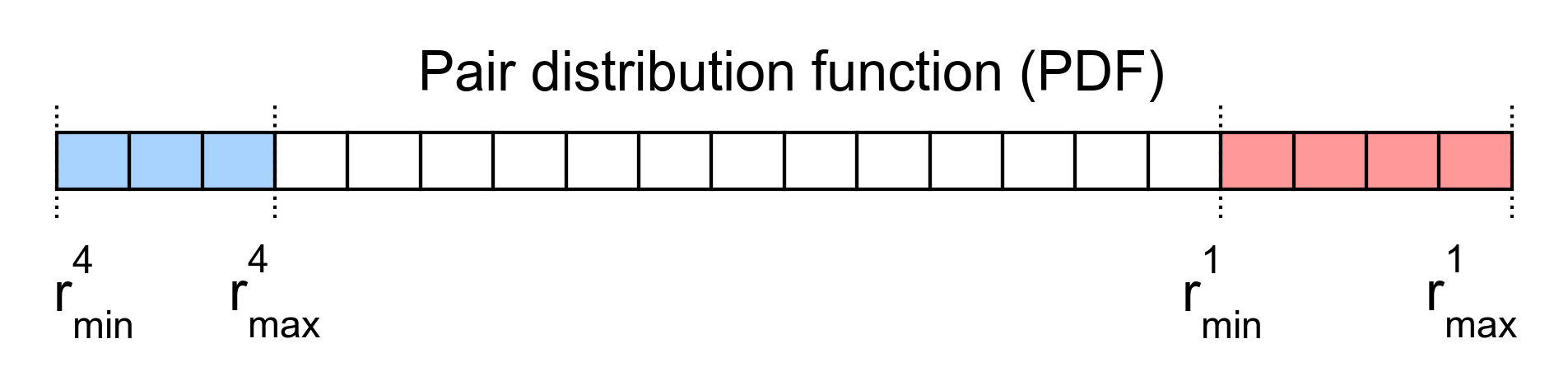}
    \caption{
        Schematic of the cell-list-based domain decomposition in a 2D toy system.  
        \textit{Top}: Representative cell pairs are color coded and grouped by center-to-center distance; the minimum ($r_\text{min}$) and maximum ($r_\text{max}$) possible interatomic distances within a pair are indicated.  
        \textit{Bottom}: Memory access pattern during PDF computation.
        Grouping cell pairs with similar center-to-center distances improves cache reuse and localizes updates to nearby PDF bin.
    }
    \label{fig:cell}
\end{figure}

We then construct a grid of cell centers and compute center-to-center distances for all unique cell pairs, storing these distances together with the corresponding pair indices.  
The resulting list is sorted by distance, as illustrated in \cref{fig:cell}.

The PDF is computed by iterating over the sorted cell pair list and accumulating contributions from all atomic pairs associated with each cell pair.  
Because the list is ordered by center-to-center distance, successive iterations tend to update nearby histogram regions, which improves cache reuse and reduces memory latency.  
This strategy is particularly effective for large or disordered systems, where otherwise random bin updates would dominate runtime.

\subsection{Parallelization}

\begin{figure}
    \centering
    \includegraphics[width=0.5\columnwidth]{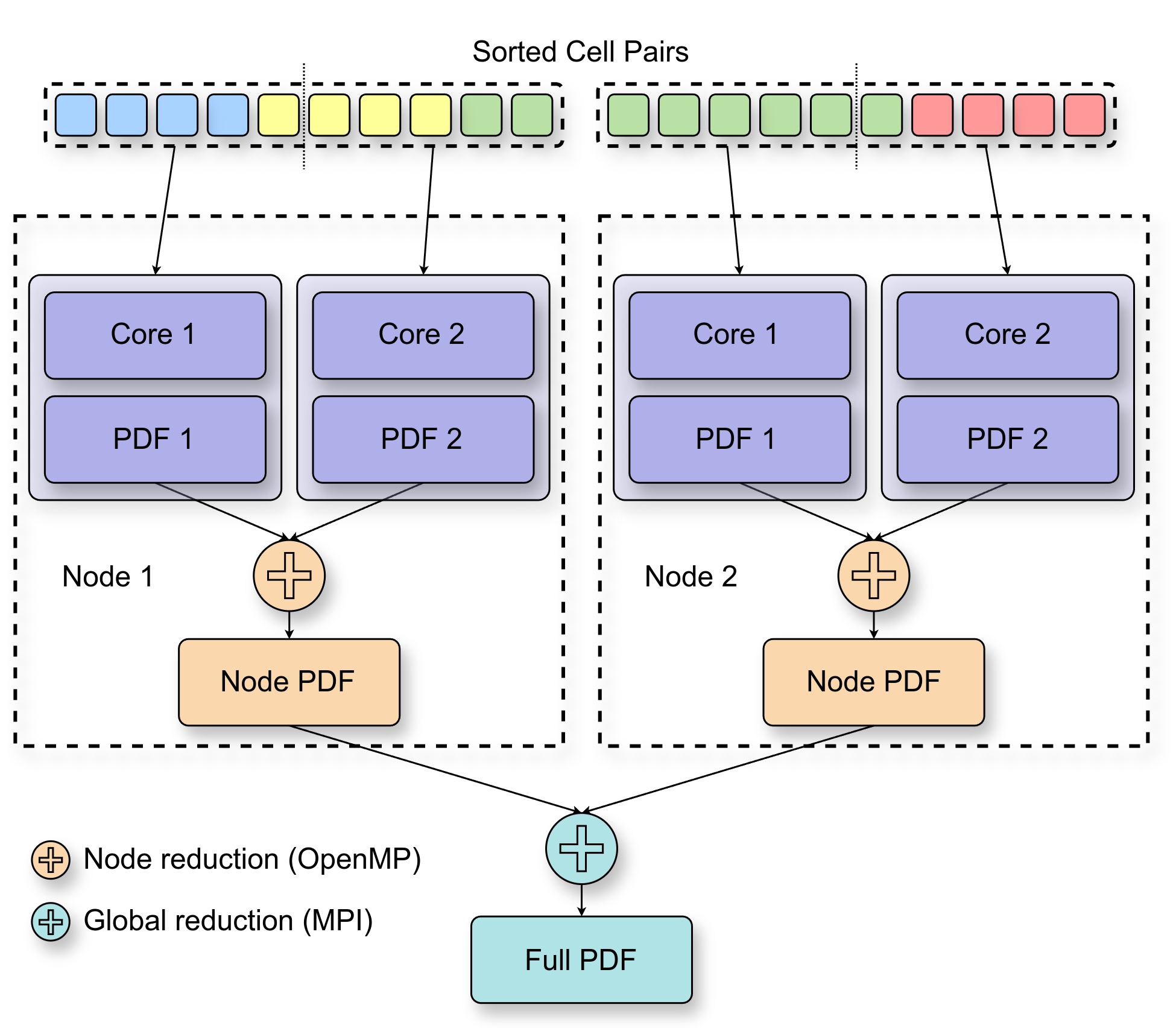}
    \caption{CPU parallelization scheme using a hybrid MPI+OpenMP model.  
    Each node runs a single MPI process that spawns multiple OpenMP threads (one per core in this example).  
    This schematic shows two MPI processes, each with two OpenMP threads.}
    \label{fig:parallelization}
\end{figure}

We accelerate PDF construction through parallelization (\cref{fig:parallelization}), using both CPU and GPU backends to support a wide range of hardware platforms.  
Although the library is designed for supercomputing clusters, it also runs efficiently on conventional workstations, desktops, and laptops.

\subsubsection{Shared memory (OpenMP)}

On CPUs, we parallelize the loop over the sorted cell pair list.
Each thread accumulates PDF contributions from its assigned cell pairs into a private buffer, thereby avoiding race conditions; the thread-local PDFs are combined in a final bin-wise reduction.
OpenMP is used for portability and compatibility with major C++ compilers.
The core loop for CPU-based PDF construction is shown in \cref{algorithm1}.

\begin{algorithm}
    \caption{CPU-based PDF computation using cell pairs}    \begin{algorithmic}
        \State $N_\text{cell-pairs} \gets \text{number of cell pairs}$
        \For{$m, n \leq N_\text{cell-pairs}$} \Comment{Parallel loop}
        \State $N_a \gets \text{number of atoms of species $a$ in cell $m$}$
        \State $N_b \gets \text{number of atoms of species $b$ in cell $n$}$
        \State $n_\text{bins} \gets \text{number of bins in the PDF}$
        \State $r_\text{max} \gets \text{maximum pair distance}$
        \For{$i \leq N_a$}
        \For{$j \leq N_b$}
        \State $\delta\mathbf{r} = \mathbf{r}_j - \mathbf{r}_i$
        \State $k = \left\lfloor \|\delta\mathbf{r}\| \, n_\text{bins} / r_\text{max} \right\rfloor$
        \State $N_k \mathrel{+}= 1$
        \State $\psi_k \mathrel{+}= (\delta\mathbf{r} \cdot \delta\mathbf{r} - \nu_k^2)$
        \EndFor
        \EndFor
        \EndFor \Comment{Reduce all thread-local PDFs}
    \end{algorithmic}\label{algorithm1}
\end{algorithm}

\subsubsection{GPU acceleration (CUDA)}

GPU parallelization is implemented using custom CUDA kernels.  
Kernel launch parameters are selected to maximize occupancy while respecting register and shared memory limits.
Each thread evaluates a pairwise distance and updates the global PDF structure in device memory.  
In contrast to the CPU implementation, a private per-thread buffer is impractical on GPUs because of the very large number of concurrent threads; therefore synchronization is handled through atomic updates to shared global memory locations.

Alongside a baseline kernel in which each thread performs both bin counting and $\psi$ accumulation, we implement a coalesced-access variant.  
In this variant, two consecutive threads operate on the same atomic pair: one updates the bin count and the other updates $\psi$.  
This organization improves memory throughput and delivers better performance across different GPU architectures.

The intensity evaluation is likewise parallelized on both CPUs and GPUs.
On CPUs, reductions are used to avoid atomic writes.
On GPUs, the work is distributed over the intensity array, with each thread computing a single intensity value by iterating over all PDF bins.
This removes the need for atomic operations in the reciprocal space evaluation.

\subsubsection{Distributed memory (MPI)}

To enable execution across multiple nodes or GPUs in a supercomputing environment (\cref{tab:parallelization}), we implement distributed memory parallelism with the Message Passing Interface (MPI).  
The sorted cell pair list is divided among MPI processes, each of which executes its assigned workload independently while OpenMP threads provide additional shared memory parallelsm.
After thread-local reduction, a global MPI reduction merges the partial PDFs into the final result.  
Although this collective communication introduces overhead, it is negligible for the large systems targeted here.

\begin{table}
    \centering
    \begin{tabular}{|c|c|}
        \hline
        \textbf{Hardware} & \textbf{Parallelization Strategy} \\
        \hline \hline
        Single CPU & OpenMP \\
        \hline
        Multiple CPUs & MPI + OpenMP \\
        \hline
        Single GPU & CUDA \\
        \hline
        Multiple GPUs & MPI + CUDA \\
        \hline
    \end{tabular}
    \caption{Parallelization strategies supported for different hardware configurations.}
    \label{tab:parallelization}
\end{table}

For multi-GPU configurations, we assign one MPI process to each GPU.
Since the current GPU implementation does not use cell lists for workload decomposition, the outer loop of the pair distance computation is partitioned across MPI processes instead.
Each GPU evaluates its assigned segment independently, and a final reduction step merges the partial PDFs into a single global result.
Further details on optimization strategies that were explored but proved ineffective are given in the supporting information.


\section{Accuracy and Numerical Validation}

\begin{figure*}[t]
     \centering
     \begin{subfigure}{0.48\textwidth}
         \raggedright\textbf{(a)}\par
         \includegraphics[width=\linewidth]{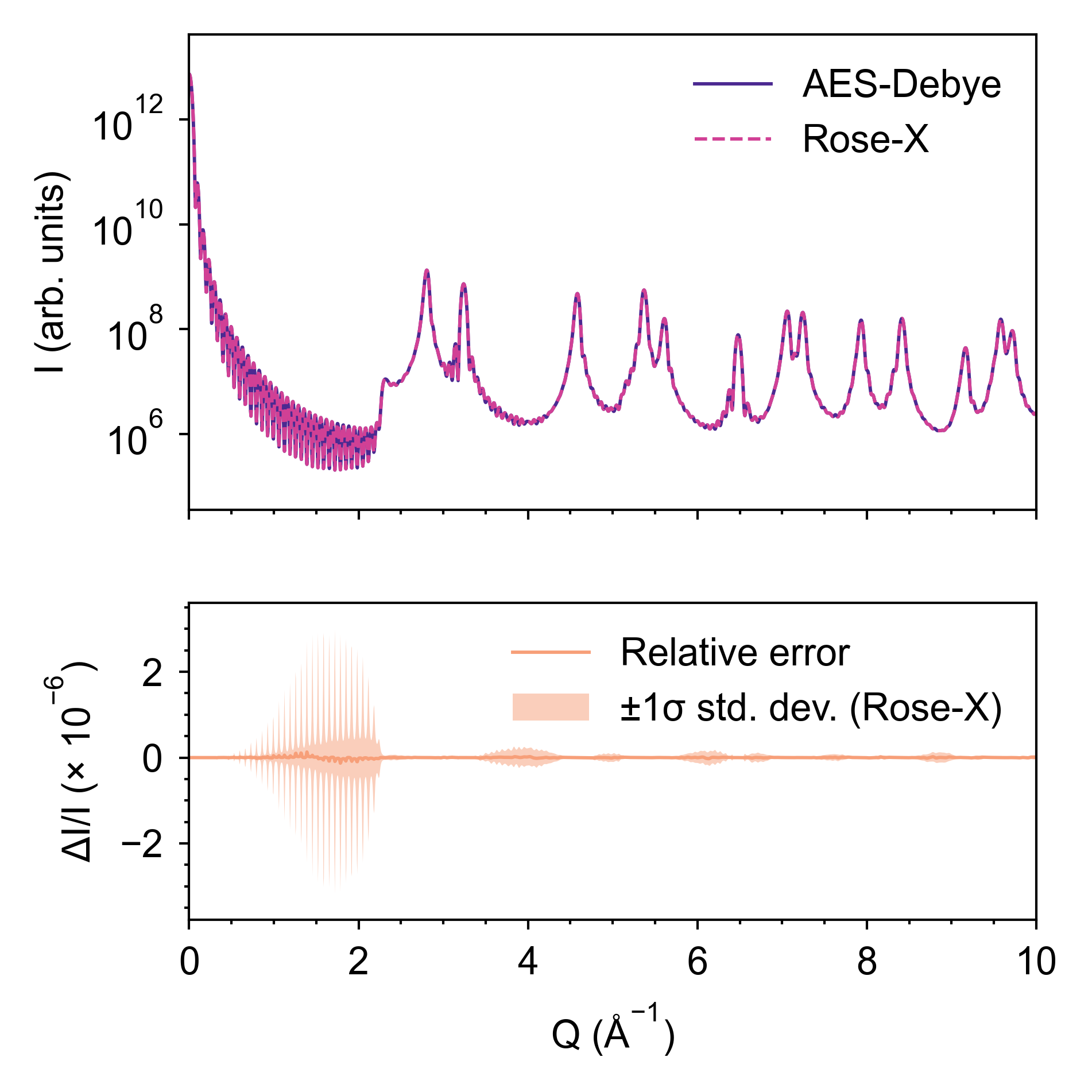}
         \label{fig:accuracy_brute}
     \end{subfigure}
     \hfill
     \begin{subfigure}{0.48\textwidth}
         \raggedright\textbf{(b)}\par
         \includegraphics[width=\linewidth]{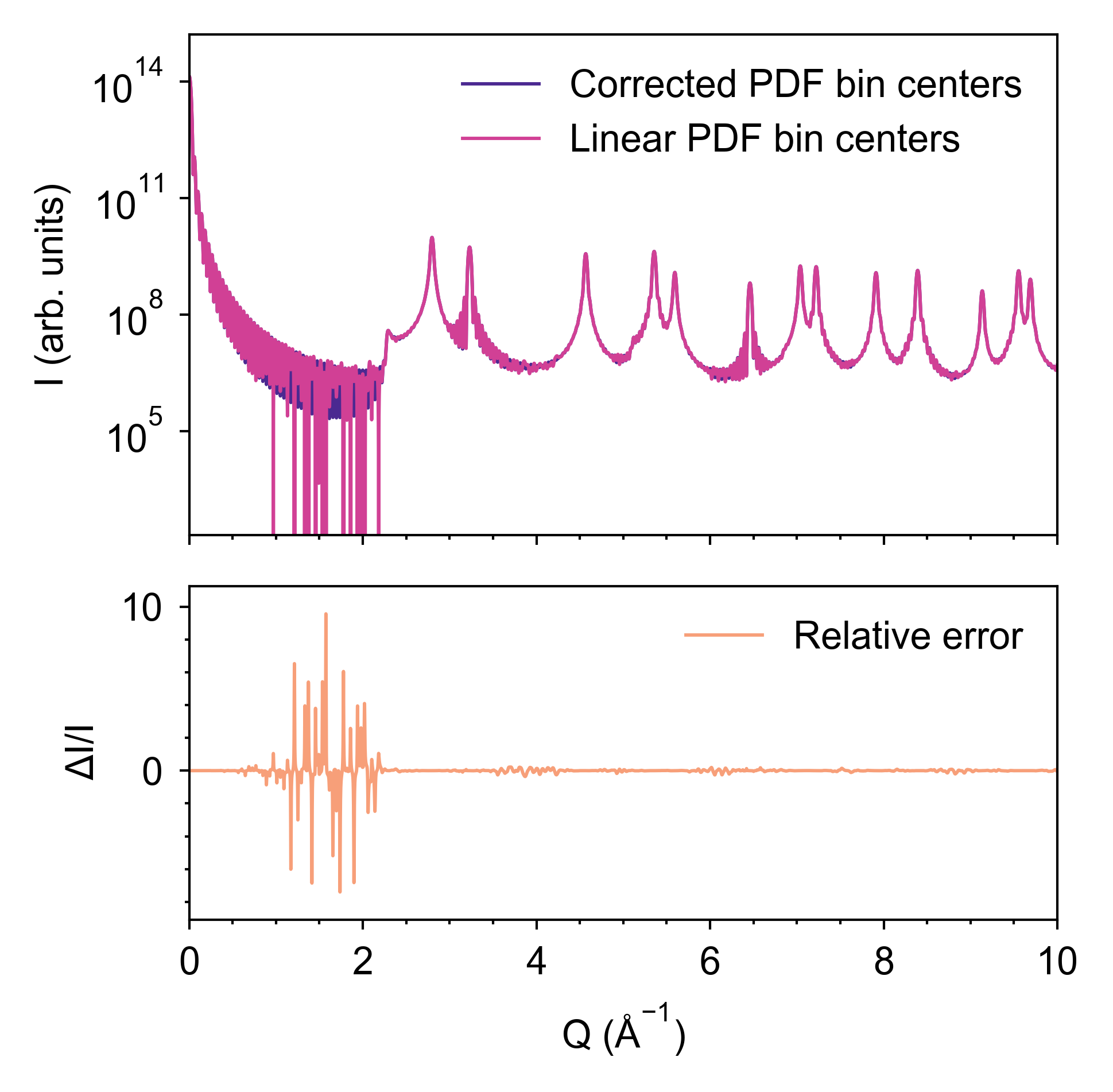}
         \label{fig:accuracy_correction}
     \end{subfigure}
     \caption{
    Validation of the PDF-based implementation and bin center correction.
    (a) Comparison of the intensity profile computed with Rose-X and with the present implementation for a Pd nanocube (side length 9.3\,nm). 
    The lower panel shows the relative error between the two profiles, together with the standard deviation of the intensity error estimated from Rose-X. 
    (b) Powder diffraction profiles computed with and without bin center correction for a crystalline Pd nanocube (side length 15.1\,nm). 
    The correction removes unphysical negative intensities and improves the diffuse background.
}
     \label{fig:accuracy}
\end{figure*}
Establishing a reliable reference for accuracy validation is challenging for large-scale Debye scattering calculations. While direct brute-force evaluation of the Debye sum is formally exact, for the system sizes considered here it can be affected by significant numerical summation errors, particularly at small scattering vectors where cancellation effects become severe. To assess the accuracy of the PDF-based implementation, we therefore compared the computed intensity profiles against those obtained with Rose-X~\cite{rose_x} (\cref{fig:accuracy}a). Although Rose-X employs approximations, it provides quantitative estimates of its calculation error, making it a suitable high-accuracy benchmark for large-scale validation.
The two profiles are nearly indistinguishable, with a maximum relative error of $\sim 1.3 \times 10^{-7}$. 
This is well within the error limits calculated using Rose-X through full error propagation.
This confirms that AES-Debye reproduces the numerical accuracy of the Rose-X formulation while enabling the architectural and performance improvements described above.

To illustrate the role of bin center correction, we computed the powder diffraction profile of a crystalline test system using two versions of the PDF: one with corrected bin centers and one without.
As shown in \cref{fig:accuracy}b, the corrected profile eliminates unphysical intensities, whereas the uncorrected profile exhibits several.
More importantly, the two profiles diverge in the diffused background region between the peaks.
Because diffuse scattering carries information about disorder, inaccuracies in this regime can bias the interpretation of both crystalline and disordered materials.
Bin center correction is therefore not merely a cosmetic improvement but an essential ingredient for high accuracy total scattering calculations.


\section{Performance benchmarks}

We benchmarked AES-Debye on cubic Pd FCC crystals with edge lengths from 4 to 40\,nm, corresponding to systems ranging from 4{,}000 to 4 million atoms.
All runs used the same $Q$ grid, PDF bin width $\Delta$, and incident wavelength to ensure direct comparability.
CPU throughput is reported in million pair distances per second per core (MPd\,s$^{-1}$\,core$^{-1}$), and GPU throughout in billion pair distances per second (BPd\,s$^{-1}$).

\begin{figure}
    \centering
    \includegraphics[width=0.5\columnwidth]{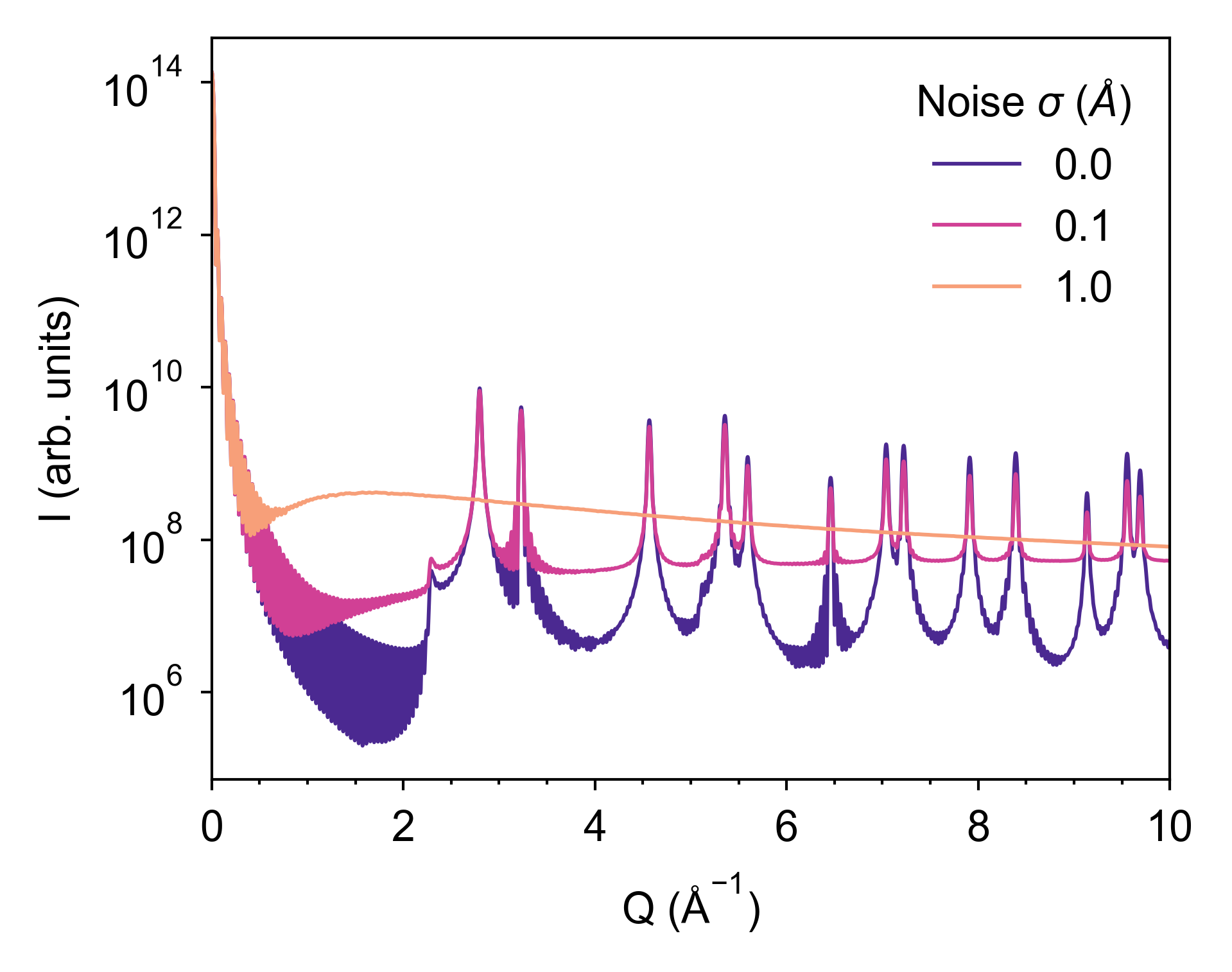}
    \caption{Diffraction profiles for a Pd nanocube (side length 10\,nm) with increasing Gaussian displacement noise $\sigma$, used to mimic thermal disorder.
    Increasing $\sigma$ broadens and attenuates the Bragg peaks, consistent with increasing structural disorder.
     The high-frequency oscillations in the background are expected finite size features associated with the parallel facets of the cube and monodisperse size, rather than numerical artifacts.} 
    \label{fig:noise_comparison}
\end{figure}

To probe sensitivity to structural disorder, we use the Debye--Einstein model~\cite{vaccari2006einstein} by adding Gaussian displacements with standard deviation $\sigma$ to the ideal lattice positions, varying $\sigma$ from 0\,\AA{} (perfect crystal) to 1\,\AA{}.
While this approach adds uncorrelated noise directly to perfectly crystalline systems, we verified that introducing structural disorder via Molecular Dynamics simulations which captures short-range correlations, produces comparable effects.
This provides a controlled spectrum of workloads, from cache-friendly crystalline systems to latency-dominated disordered ones, and is used throughout the scaling studies. 
The corresponding evolution of the diffraction profile is shown in \cref{fig:noise_comparison}.

\subsection{CPU scaling}

CPU benchmarks were performed on a single node of the Fritz cluster at the Erlangen National High Performance Computing Center (NHR@FAU)~\cite{fauFritzNHRFAU}, using dual socket Intel Xeon Platinum 8360Y processors (36 cores per socket, 54\,MB L3/socket, SMT disabled).  
All runs were executed at a fixed frequency of 2.2\,GHz to eliminate variability from turbo boost and dynamic frequency scaling.
The code was compiled with Intel oneAPI C++ (2023).

\begin{figure*}[t]
    \centering
    \begin{subfigure}{0.48\textwidth}
        \raggedright\textbf{(a)}\par
        \includegraphics[width=\linewidth]{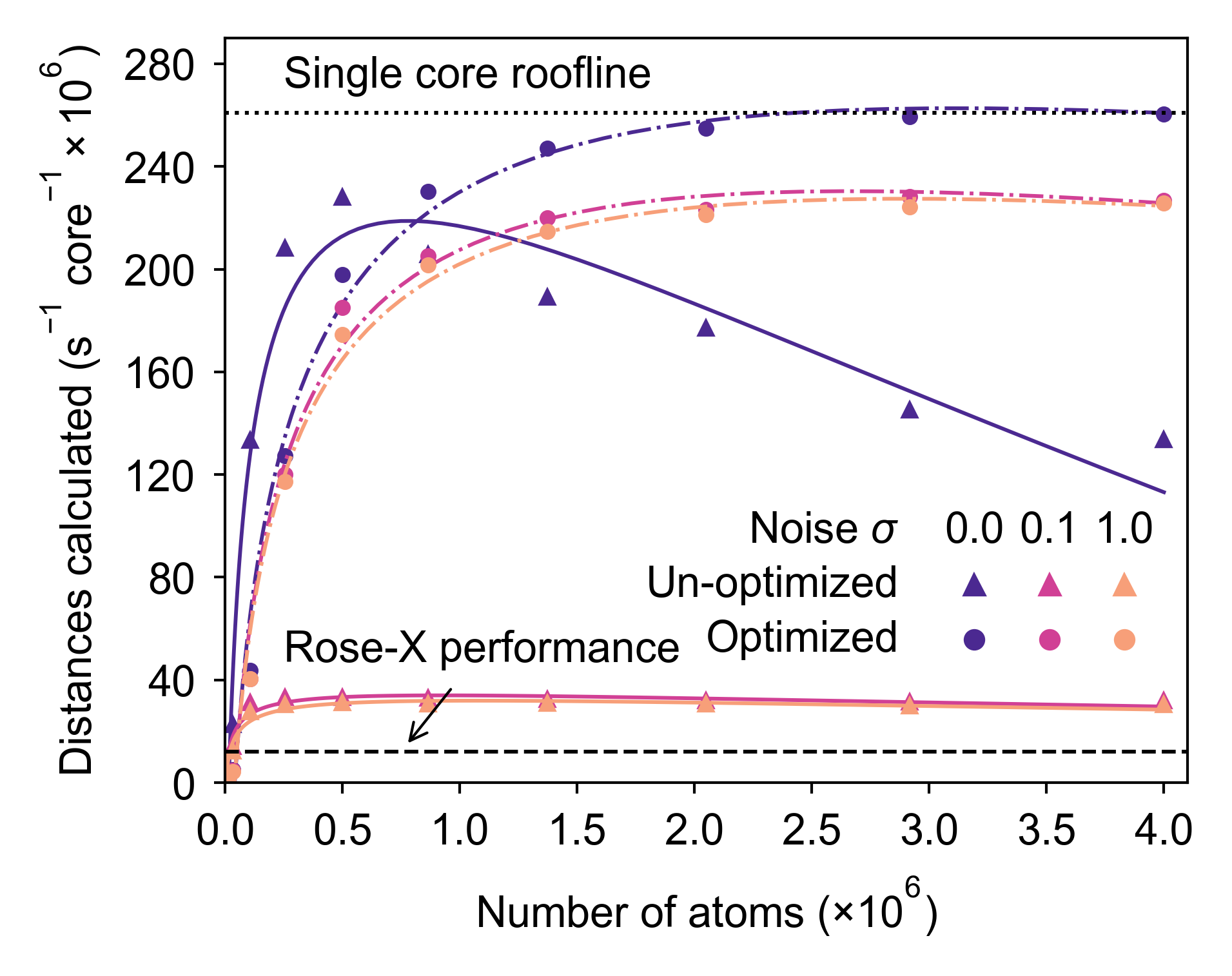}
        \label{fig:atomic_scaling}
    \end{subfigure}
    \hfill
    \begin{subfigure}{0.48\textwidth}
        \raggedright\textbf{(b)}\par
        \includegraphics[width=\linewidth]{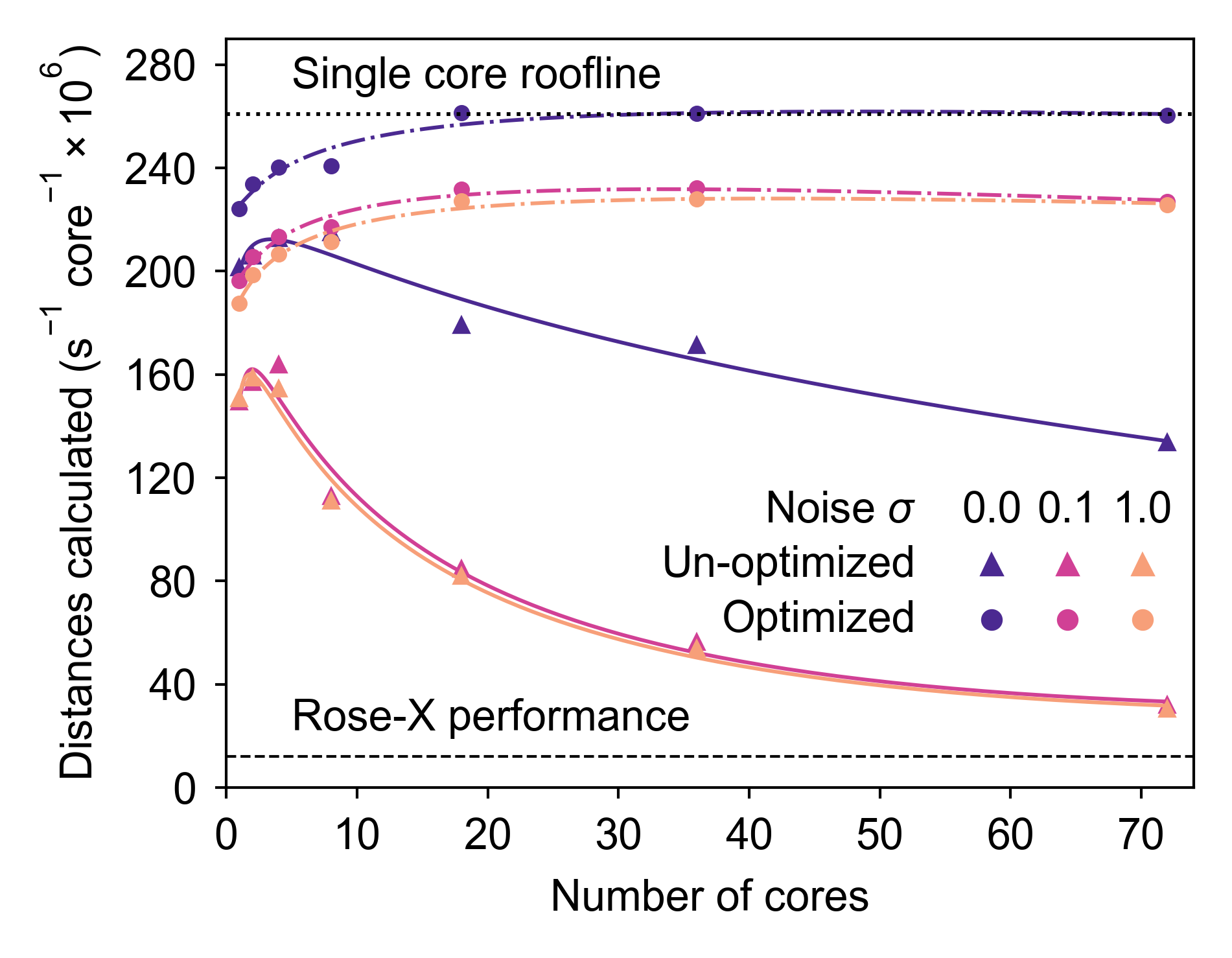}
        \label{fig:scaling}
    \end{subfigure}

    \vspace{1em} 

    \begin{subfigure}{0.48\textwidth}
        \raggedright\textbf{(c)}\par
        \includegraphics[width=\linewidth]{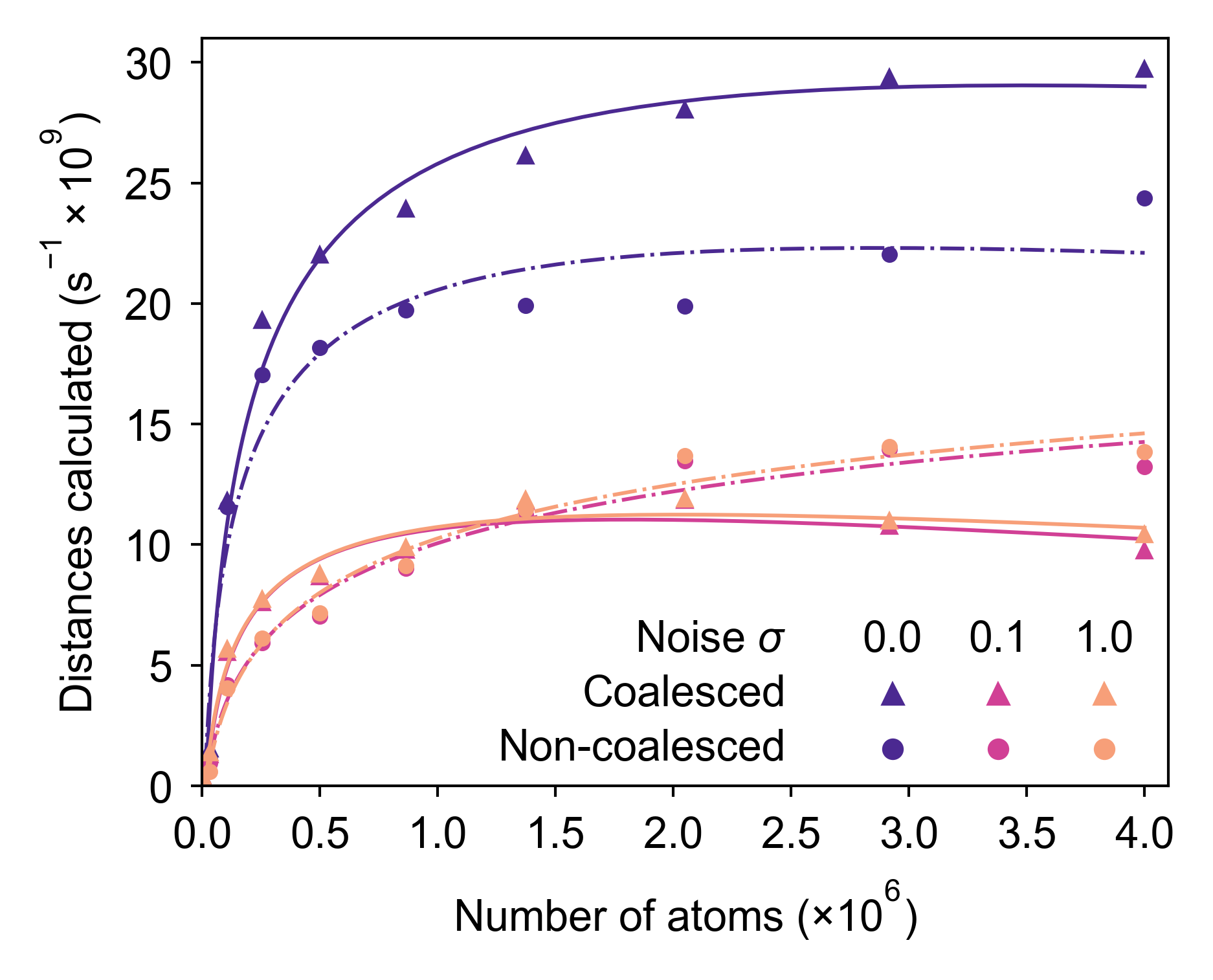}
        \label{fig:a40}
    \end{subfigure}
    \hfill
    \begin{subfigure}{0.48\textwidth}
        \raggedright\textbf{(d)}\par
        \includegraphics[width=\linewidth]{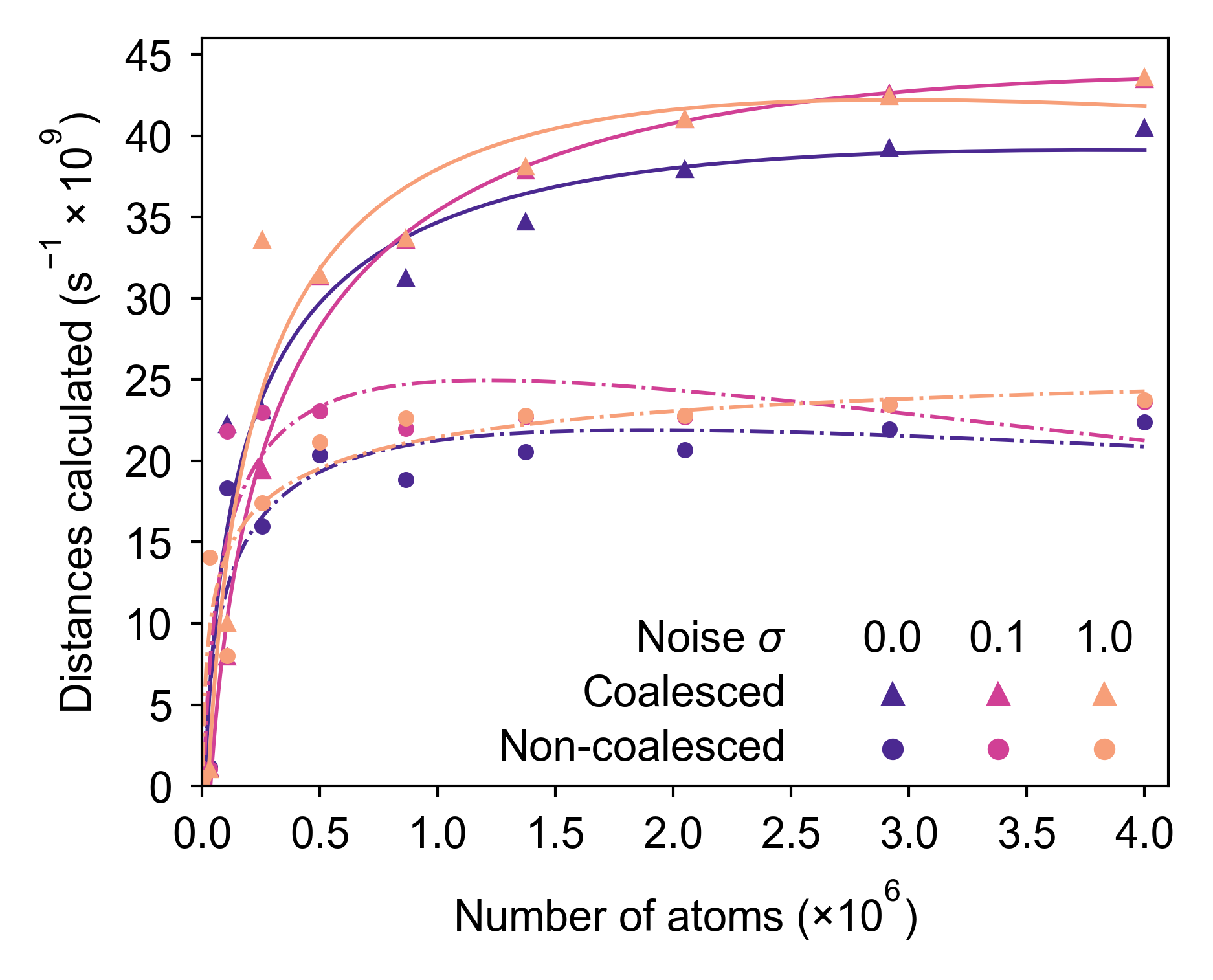}
        \label{fig:a100}
    \end{subfigure}

    \caption{
Performance scaling across CPUs and GPUs for different disorder levels $\sigma$.
(a) Throughput versus atom count on 72 CPU cores; the dashed lines indicate the single-core roofline and a comparison to the existing earlier implementation in Rose-X.
(b) Strong scaling on a single CPU node.
(c) Asymptotic performance on an NVIDIA A40 (48\,GB GDDR6, 6\,MB L2 cache).  
(d) Asymptotic performance on an NVIDIA A100 (40\,GB HBM2e, 25\,MB L2 cache).  
}
    \label{fig:all_scaling}
\end{figure*}

\cref{fig:all_scaling}a shows the per-core PDF throughput as a function of system size.  
For the baseline implementation without cell lists, crystalline systems reach a maximum near 500{,}000 atoms and then decline, whereas disordered systems saturate earlier at substantially lower throughput.
This drop reflects increasing cache pressure as more PDF bins are updated and the working set spills beyond L3 cache into main memory.

The cell list implementation largly removes this large-size degradation.
Its throughput increases with system size, surpasses the baseline at approximately 750{,}000 atoms, and approaches the single-core roofline for large crystalline structures.
For both crystalline and disordered systems, the cell list implementation overtakes the baseline and sustains higher throughput across the full benchmark range.
The precise crossover depends on system size, cell count, disorder level, and CPU architecture.

Strong-scaling results are sown in \cref{fig:all_scaling}b.
The cell list implementation consistently outperforms the baseline and improves scaling in both crystalline and disordered cases, with speedups of up to 1.5$\times$ and 5.5$\times$, respectively.
In the baseline implementation, performance of disordered systems degrades with increasing core count because more threads contend for shared L3 cache resources.  
By localizing bin updates, the cell list implementation improves cache reuse and reduces traffic, substantially reducing the loss.

\subsection{GPU scaling}

GPU benchmarks were performed on NVIDIA A100 and A40 devices on the Alex cluster at NHR@FAU~\cite{fauAlexNHRFAU}.  
The A100 provides 40\,GB HBM2e memory with 108 SMs, while the A40 provides 48\,GB GDDR6 memory with 84 SMs.  
The GPU implementation was compiled with CUDA~12.4.

On both GPUs, throughput increases with system size and then saturates.
We also compared kernels with coalesced and non-coalesced memory access patterns.  
On the A40, crystalline systems achieve up to 30\,BPd\,s$^{-1}$ (\cref{fig:all_scaling}c), but the throughput in disordered systems fall to less than half of this value, reflecting the limited 6\,MB L2 cache and the resulting sensitivity to irregular memory access.
In the strongly disordered regime, A40 performance becomes comparable to that of a single Fritz node with 72 cores tested in the earlier section.

On the A100, throughput reaches 40\,BPd\,s$^{-1}$ for crystalline systems and 43\,billion for disordered systems (\cref{fig:all_scaling}d), highlighting the benefit of the larger 25\,MB L2 cache under irregular memory access.
Coalesced access consistently outperforms non-coalesced access in crystalline systems despite the additional arithmetic, because the gain in memory efficiency outweigh the extra work.
In disordered systems, the A100 shows similar performance for both access modes, whereas the A40 performs worse with the coalesced variant, indicating stronger cache contention on that architecture.
For this reason, both kernel variants are provided.

\subsection{Node scaling (MPI)}

\begin{figure}
    \centering
    \includegraphics[width=0.5\columnwidth]{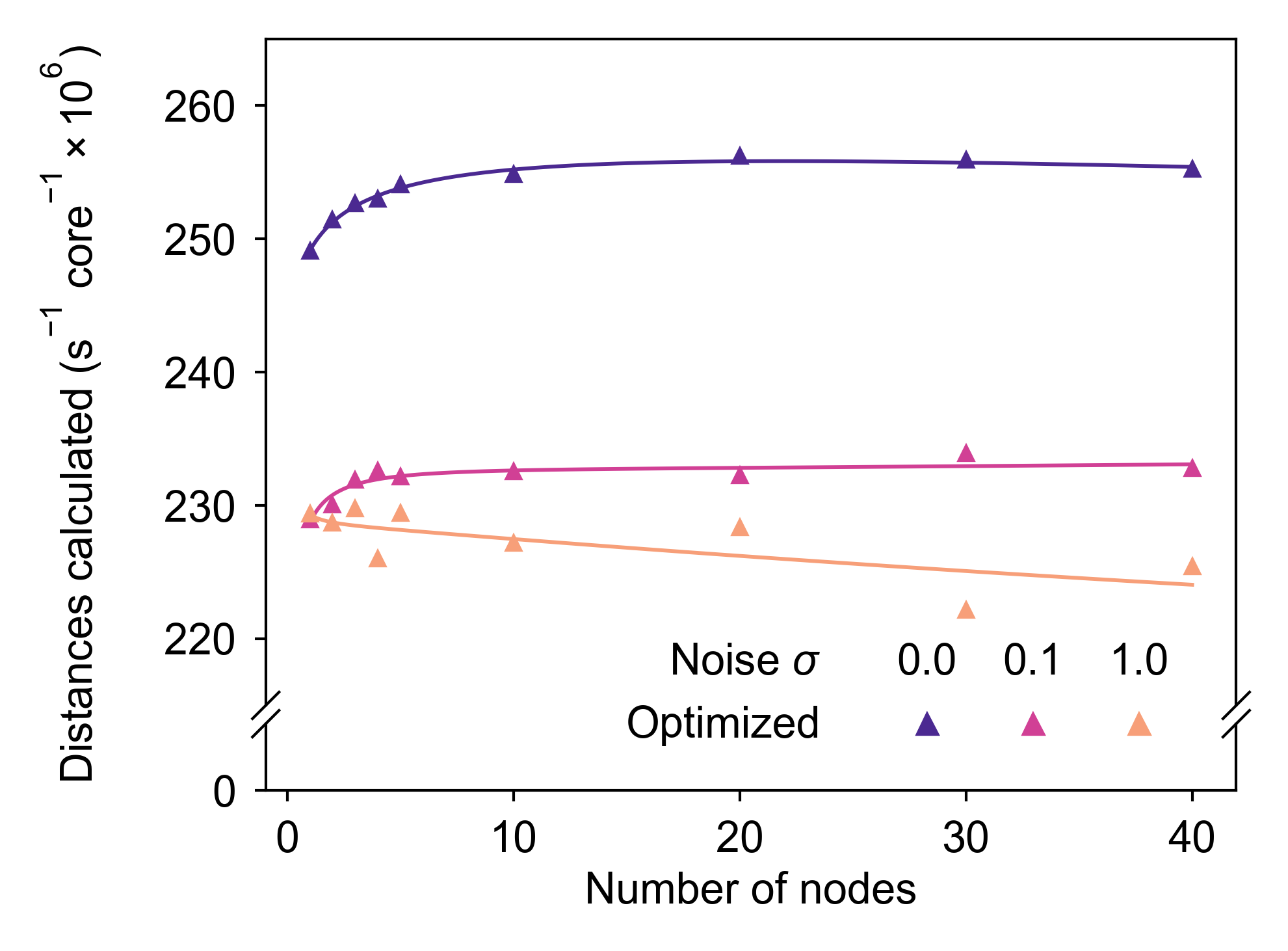}
    \caption{Strong-scaling performance as a function of node count for a 4\,million atom system. 
    Only the optimized implementation is shown. 
    The performance loss due to communication overhead remains minimal ($<5\%$) over the tested range.}
    \label{fig:mpi}
\end{figure}

To assess strong scaling across multiple nodes, we computed a diffraction profile for a large atomistic model (\cref{fig:mpi}). 
The implementation scales efficiently up to 40 nodes on the Fritz cluster because the dominant cost, pair accumulation, is distributed across MPI rank, with communication limited to a final global reduction.

For load balancing, we estimate the work associated with each cell pair from the product of the occupancies of the two cells.
The sorted cell pair list is then partitioned into contiguous chunks with approximately equal estimated cost, and these chunks are assigned to MPI ranks. 
This strategy provides good load balance even for disordered systems.
    
A modest performance loss with increasing node count is unavoidable because serial components reduction overhead become more visible as the parallel portion shortens.
The degradation is slightly more pronounced for disordered systems, reflecting residual imbalance that is difficult to eliminate fully with a simple contiguous partitioning of the cell pair list.
Nevertheless, the penalty remains below 5\% over the tested range.

For workloads consisting of many trajectory frames but only moderate system sizes, throughput is better improved through ensemble parallelism: using fewer nodes per frame and processing multiple frames concurrently. 

\subsection{Performance comparison}

Cross-code comparisons are necessarily approximate because implementations differ in numerical choices, execution models, and hardware targets.
As a representative reference, we compared AES-Debye with DebyeCalculator~\cite{Johansen_anker2024debye}, a recent brute force implementation designed primarily for GPU execution.
All benchmarks were performed on a single NVIDIA A40 GPU and a single AMD EPYC 7713 (Milan, Zen~3) CPU.

\begin{table}
    \centering
    \begin{tabular}{|l|c|c|}
        \hline
        \textbf{Implementation} & \textbf{CPU} & \textbf{GPU} \\
        \hline\hline
        DebyeCalculator & 0.14 (128 threads) & 6.5 \\
        \hline
        AES-Debye & 371 (8 threads) & 1531 \\
        \hline
    \end{tabular}
    \caption{Throughput in million pair distances per second (MPd\,s$^{-1}$) for a Pd nanoparticle with 32{,}000 atoms and Gaussian displacements of standard deviation $\sigma=0.01$\,\AA. 
    Because DebyeCalculator does not support substantially large systems, the comparison is performed near the upper end of the problem sizes reported in~\cite{Johansen_anker2024debye}.}
    \label{tab:performance}
\end{table}

As summarized in \cref{tab:performance}, AES-Debye achieves substantially higher throughout than DebyeCalculator on this benchmark.
On the A40 GPU, AES-Debye reaches 1531~MPd\,s$^{-1}$, compared with 6.5~MPd\,s$^{-1}$ for DebyeCalculator, corresponding to a speedup of approximately \(235\times\).
On the CPU, AES-Debye reaches 371~MPd\,s$^{-1}$ using 8 threads, whereas DebyeCalculator reports 0.14~MPd\,s$^{-1}$ with 128 threads. 
Although these CPU runs use different thread counts and therefore should not be interpreted as a strictly controlled scaling comparison, the absolute performance gap remains very large.
Overall, AES-Debye outperforms DebyeCalculator by a wide margin on both CPU and GPU for this problem size.


\section{Applications}

\begin{figure*}[ht]
    \centering
    \setbox0=\hbox{\includegraphics[width=0.48\textwidth]{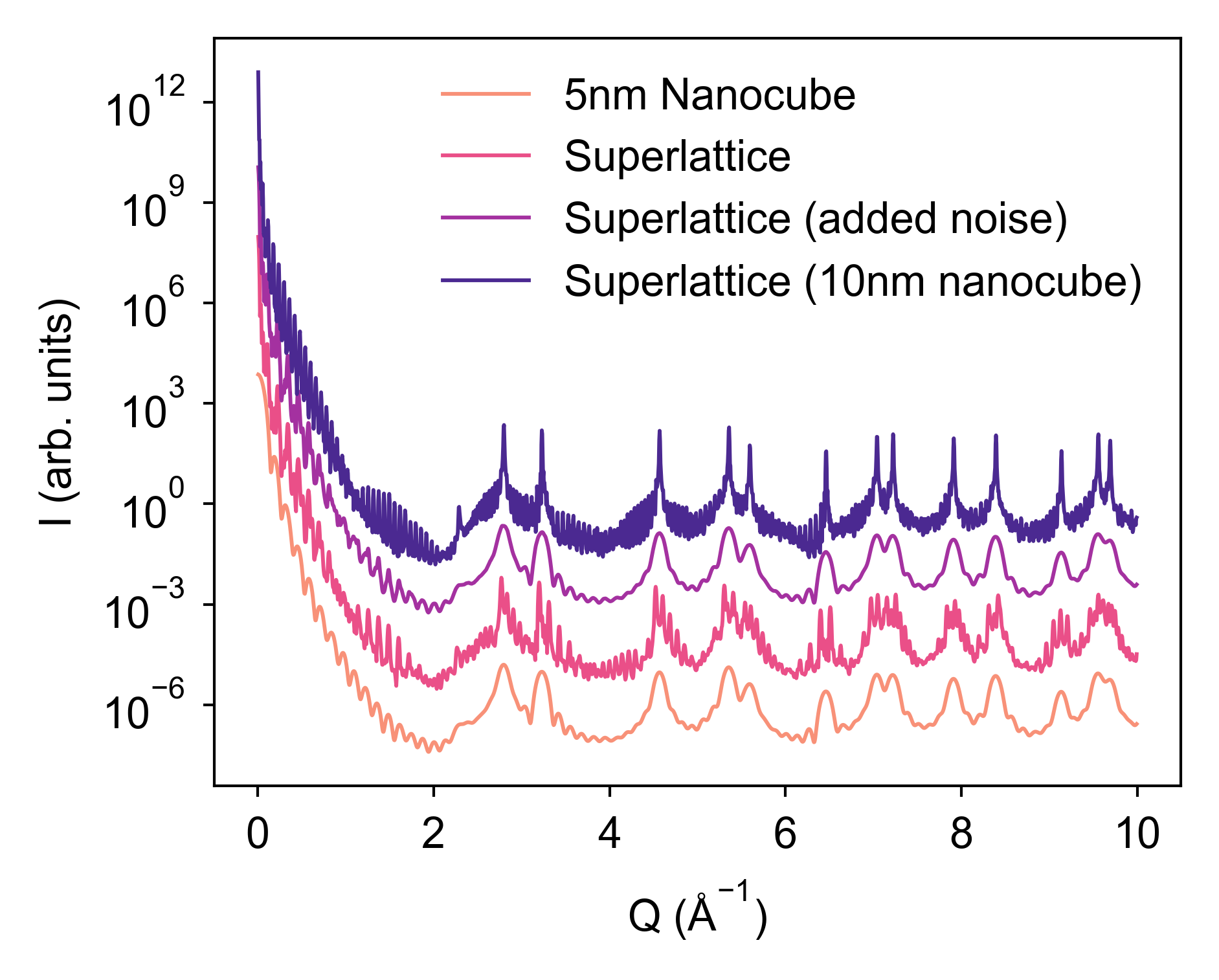}}
    \begin{subfigure}[t]{0.48\textwidth}
        \raggedright\textbf{(a)}\par\smallskip
        \begin{minipage}[b][\ht0][c]{\linewidth}
            \centering
            \includegraphics[width=\linewidth, height=\ht0, keepaspectratio]{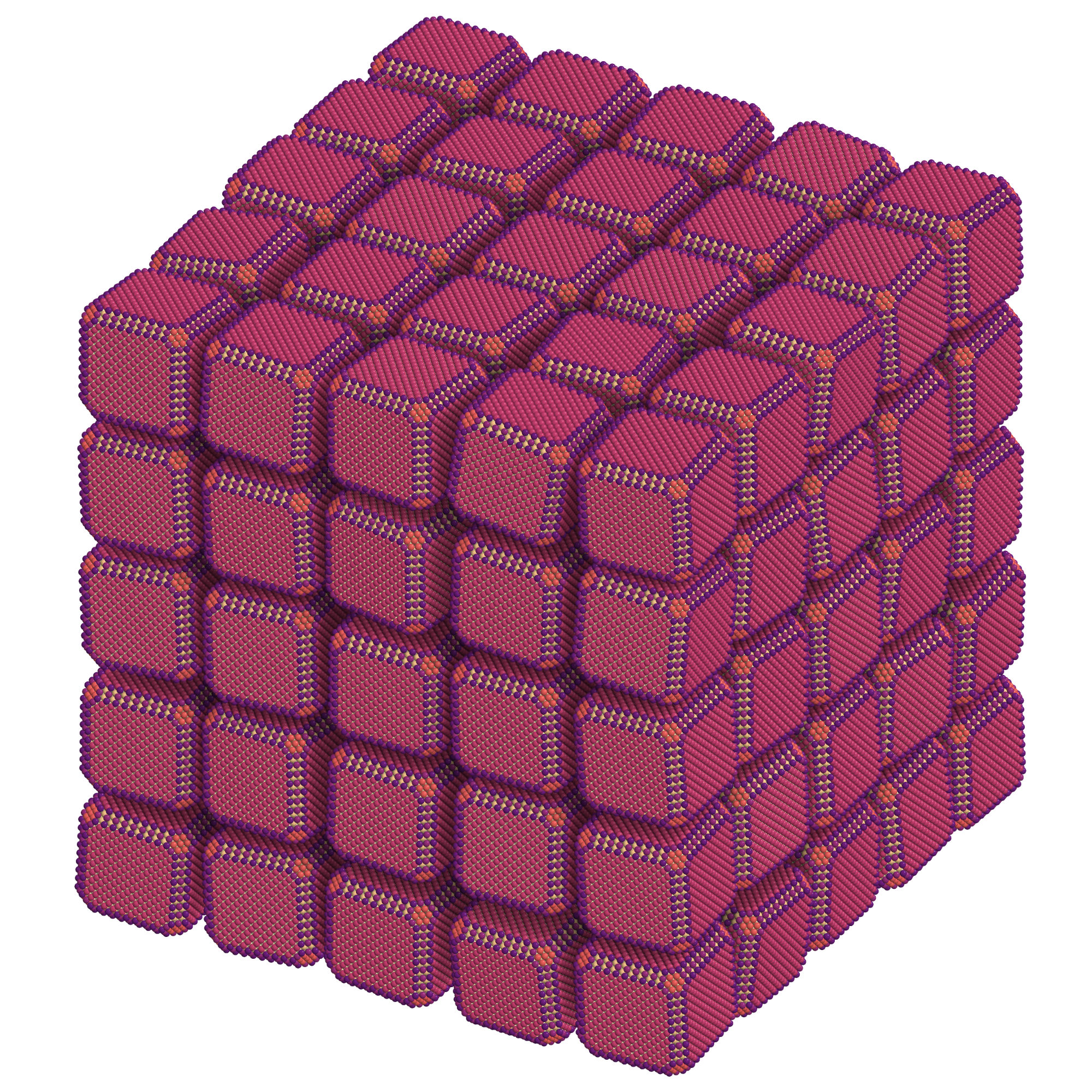}
        \end{minipage}
        \label{fig:struct_a}
    \end{subfigure}
    \hfill
    \begin{subfigure}[t]{0.48\textwidth}
        \raggedright\textbf{(b)}\par\smallskip
        \usebox0
        \label{fig:profile_a}
    \end{subfigure}
    \caption{
        (a) Pd nanocubes (5\,nm) assembled into a $5\times5\times5$ simple cubic superlattice with positional noise ($\sigma=0.05\text{nm}$).
        (b)Comparison of four diffraction profiles for nanocube systems (from bottom to top): an isolated single particle, an ideal simple cubic assembly of 5\,nm truncated nanocubes, a similar assembly with translational and rotational noise, which closely matches the single particle profile, and an ideal superlattice assembly of $\approx$10\,nm truncated nanocube building blocks.
    }
    \label{fig:assembly_combined}
\end{figure*}

\begin{figure*}[ht]
    \centering
    \setbox0=\hbox{\includegraphics[width=0.48\textwidth]{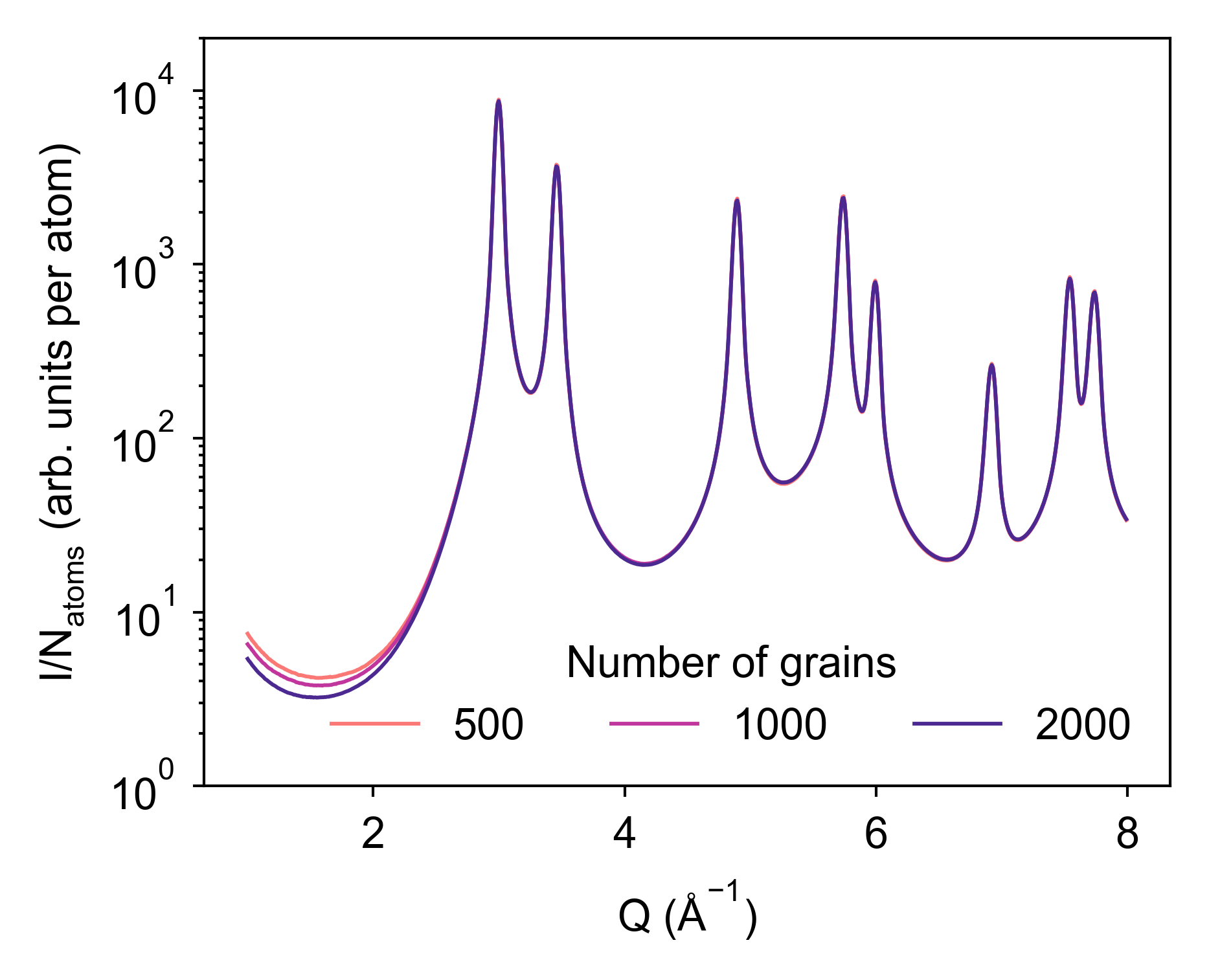}}
    \begin{subfigure}[t]{0.48\textwidth}
        \raggedright\textbf{(a)}\par\smallskip
        \begin{minipage}[b][\ht0][c]{\linewidth}
            \centering
            \includegraphics[trim=0 0 0 .5cm, clip, width=\linewidth, height=\ht0, keepaspectratio]{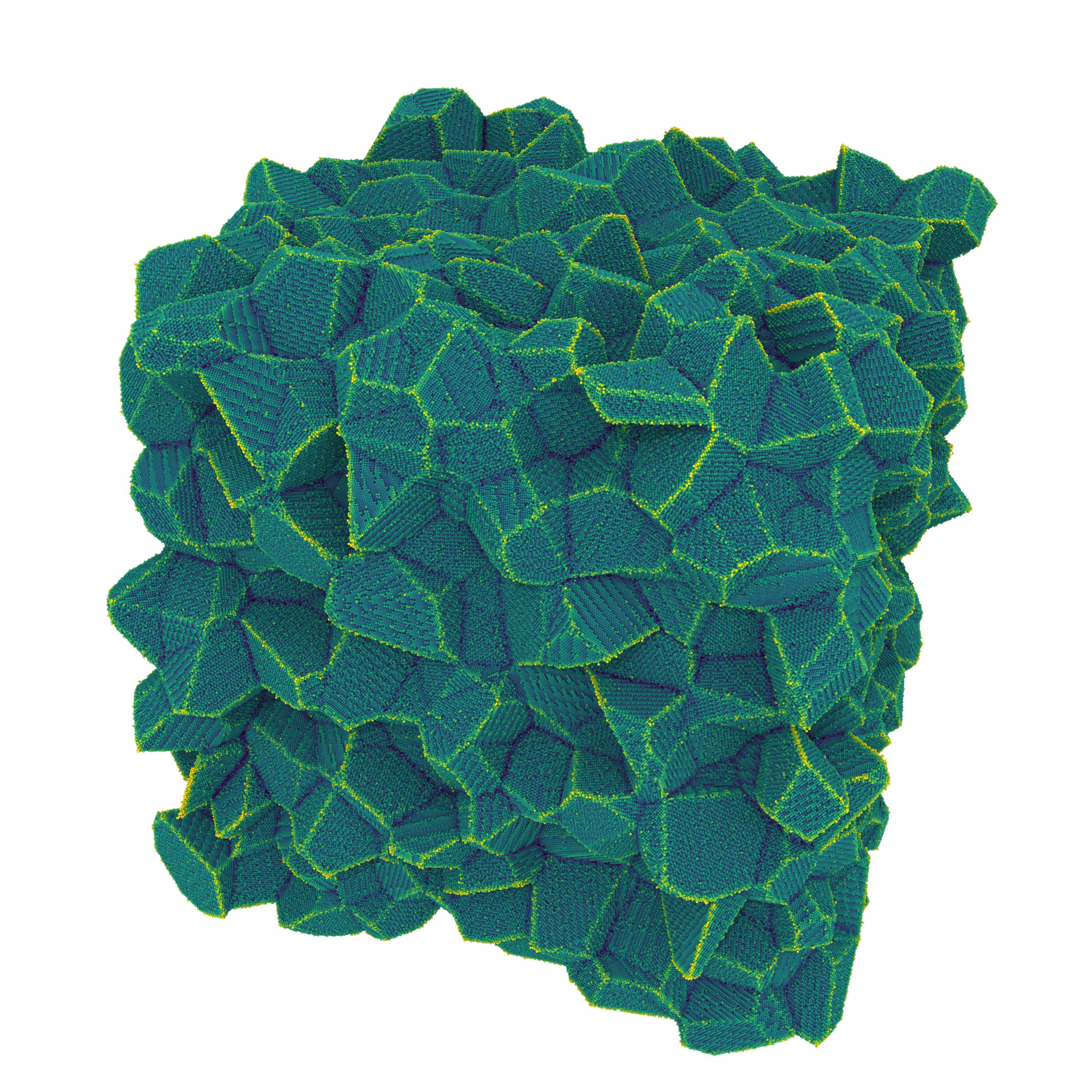}
        \end{minipage}
        \label{fig:struct_b}
    \end{subfigure}
    \hfill
    \begin{subfigure}[t]{0.48\textwidth}
        \raggedright\textbf{(b)}\par\smallskip
        \usebox0
        \label{fig:profile_b}
    \end{subfigure}
    \caption{
        (a) Polycrystalline Cu (FCC) sample with 1000 grains; atoms colored by coordination number. Atomic positions were averaged over multiple molecular dynamics frames.
        (b) Powder diffraction profiles for polycrystalline samples with different numbers of grains, normalized by atom count. The largest differences appear in the low-angle region and in the relative peak intensities.
    }
    \label{fig:poly_combined}
\end{figure*}

\begin{figure*}[ht]
    \centering
    \setbox0=\hbox{\includegraphics[width=0.48\textwidth]{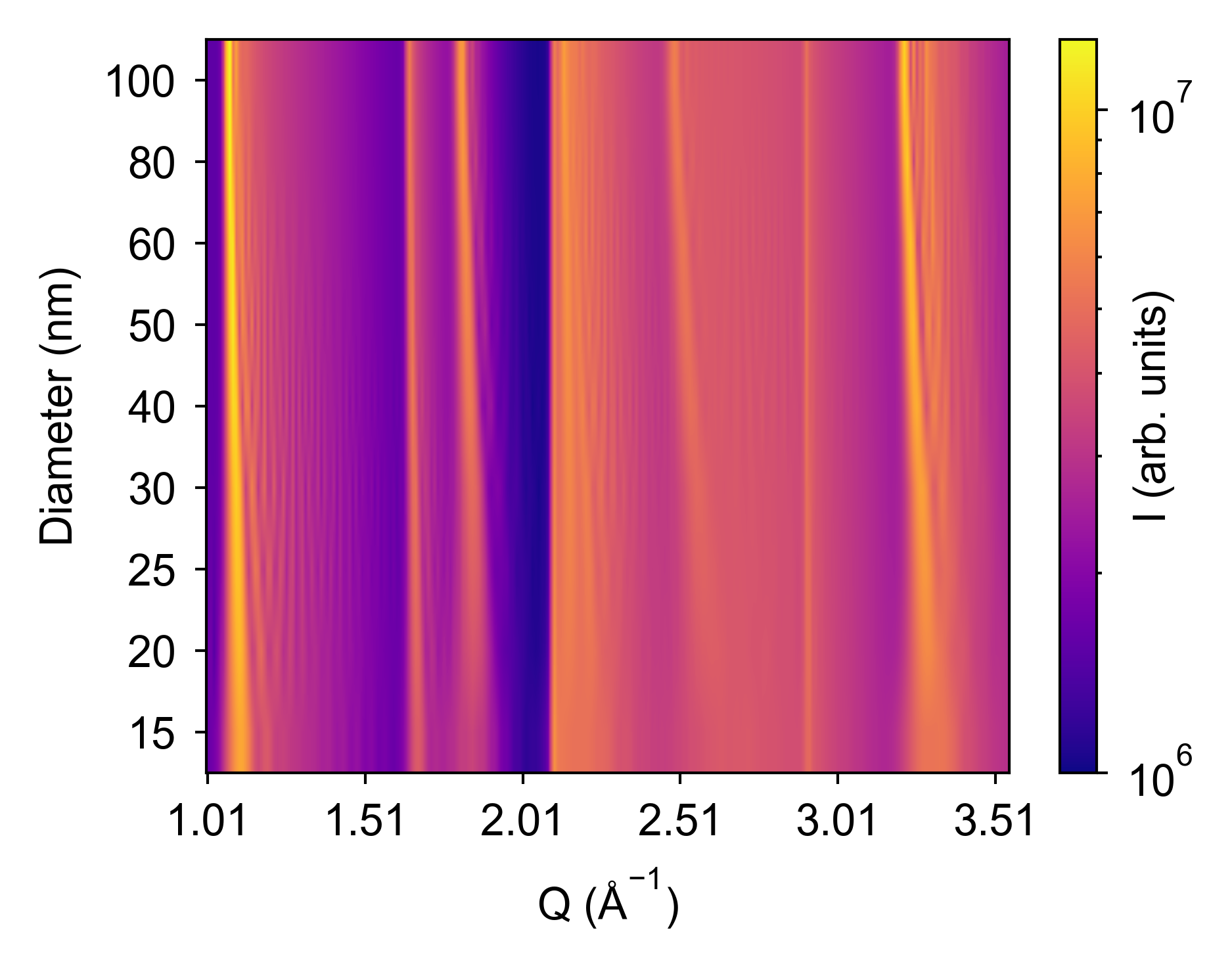}}
    \begin{subfigure}[t]{0.48\textwidth}
        \raggedright\textbf{(a)}\par\smallskip
        \begin{minipage}[b][\ht0][c]{\linewidth}
            \centering
            \includegraphics[width=\linewidth, height=\ht0, keepaspectratio]{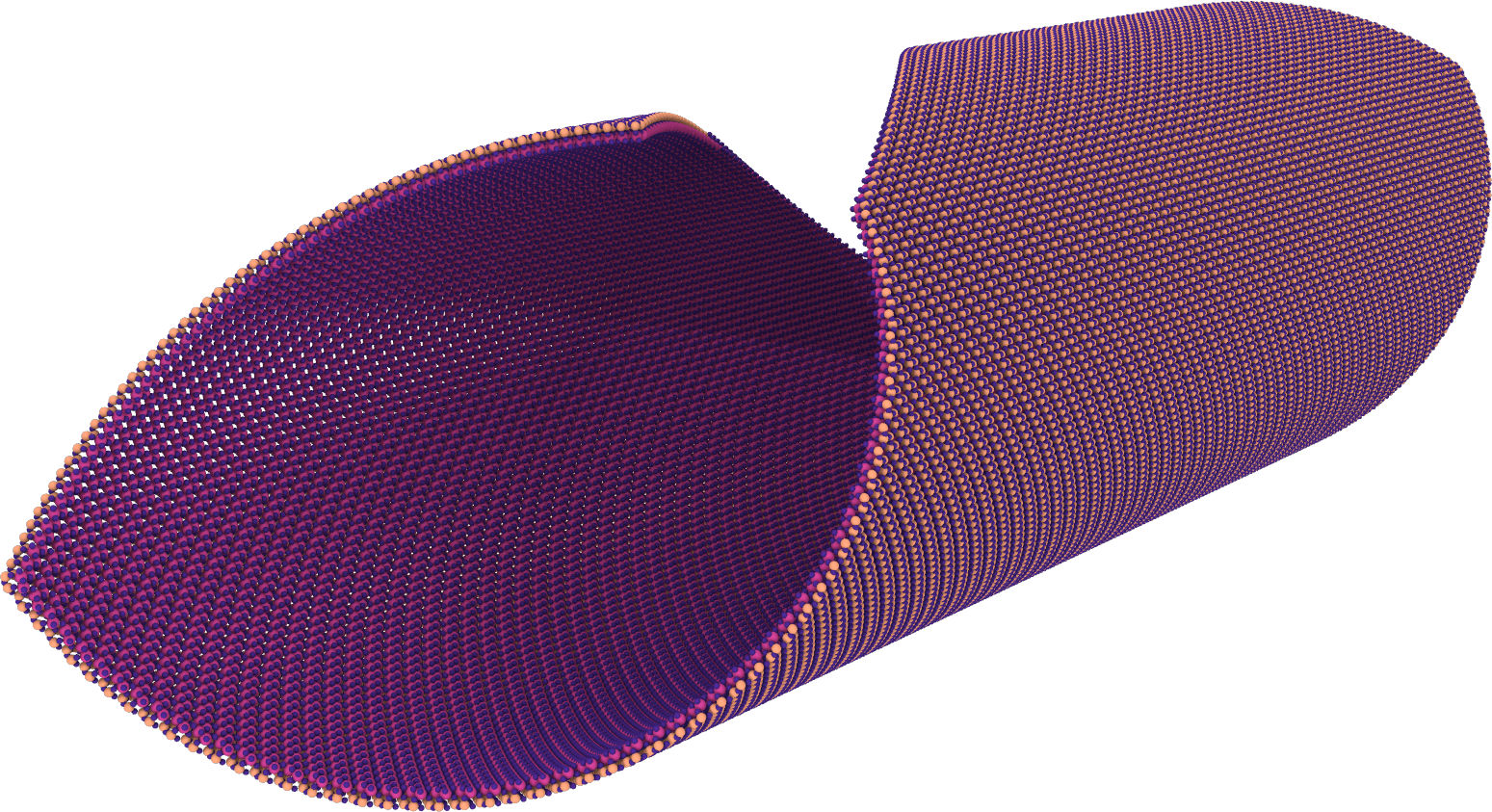}
        \end{minipage}
        \label{fig:struct_c}
    \end{subfigure}
    \hfill
    \begin{subfigure}[t]{0.48\textwidth}
        \raggedright\textbf{(b)}\par\smallskip
        \usebox0
        \label{fig:profile_d}
    \end{subfigure}
    \caption{
        (a) Atomic model of a halloysite nanotube generated from a kaolinite sheet. 
        (b) Two-dimensional visualization of diffraction intensity as a function of $Q$ and tube diameter.
    }
    \label{fig:halloysite_combined}
\end{figure*}

\begin{figure*}[ht]
    \centering
    \setbox0=\hbox{\includegraphics[width=0.48\textwidth]{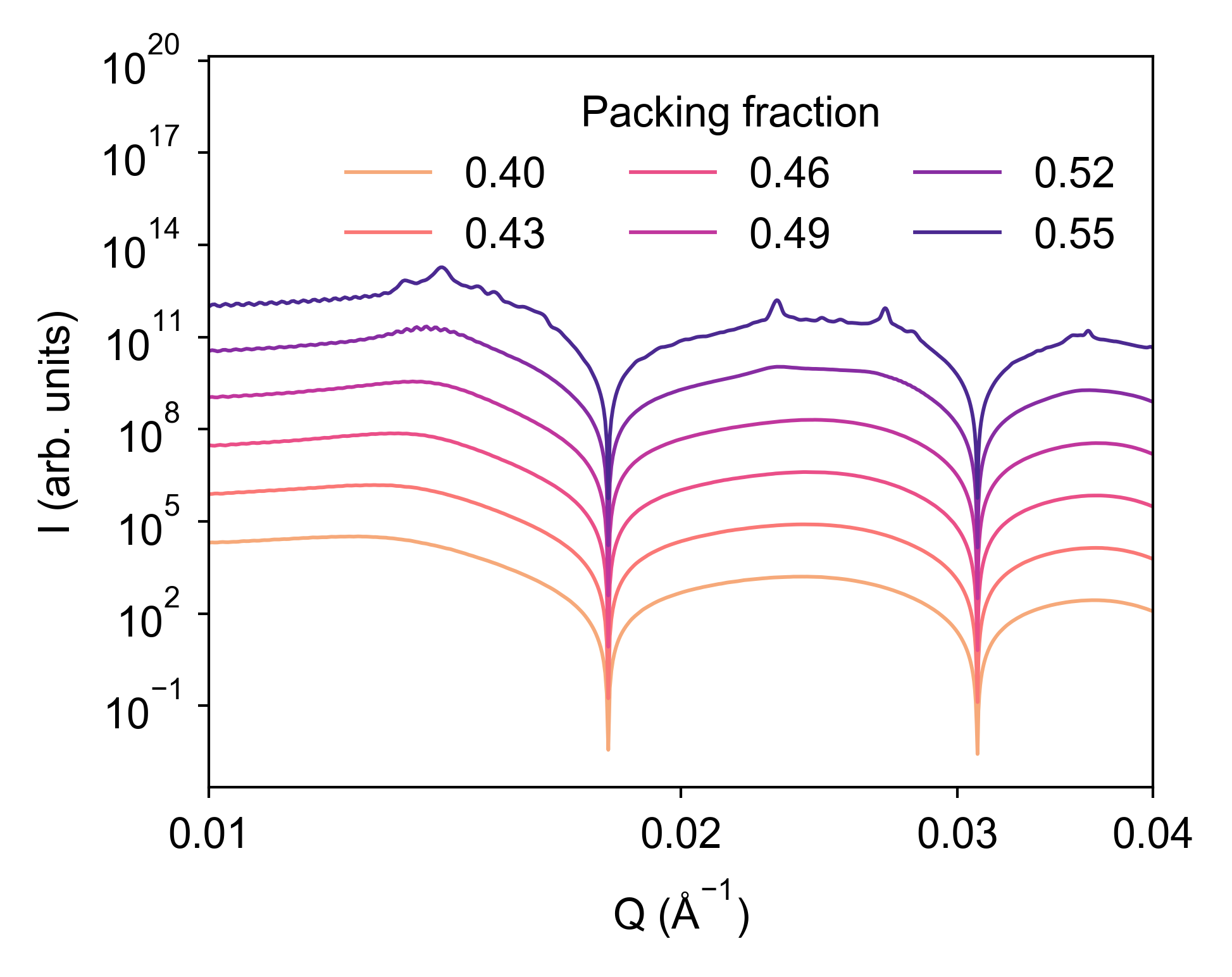}}
    \begin{subfigure}[t]{0.48\textwidth}
        \raggedright\textbf{(a)}\par\smallskip
        \begin{minipage}[b][\ht0][c]{\linewidth}
            \centering
            \includegraphics[width=\linewidth, height=\ht0, keepaspectratio]{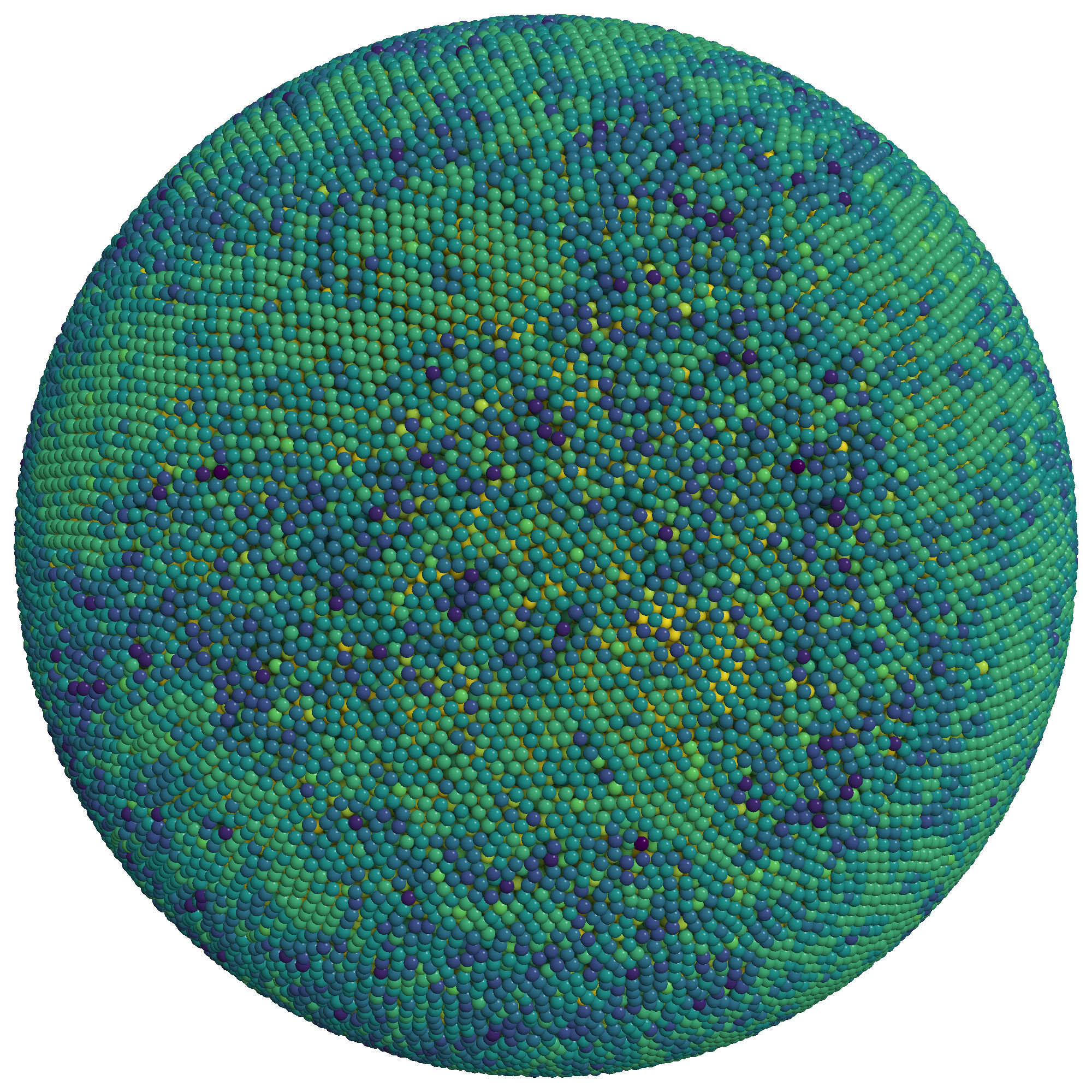}
        \end{minipage}
        \label{struct_e}
    \end{subfigure}
    \hfill
    \begin{subfigure}[t]{0.48\textwidth}
        \raggedright\textbf{(b)}\par\smallskip
        \usebox0
        \label{fig:profile_f}
    \end{subfigure}
    \caption{
        (a) Spherical nanoparticles (diameter: 200nm) crystallized in a spherical confinement, at a packing faction of 0.55. The particles are colored using their coordination number to highlight the surface defects that form due to the spherical confinment.
        (b) Small angle scattering profiles. As the packing fraction increases (bottom to top, shifted for clarity), we start to see the FCC superlattice peaks on above the sphere form factor in the background.
    }
    \label{fig:colloidal_combined}
\end{figure*}

To complement the synthetic benchmarks, we present four representative applications that showcase the scalability and efficiency of the AES-Debye implementation.
Nanocube superlattices highlight the value of the Debye equation for hierarchical self-assembly by resolving how superlattice ordering modifies the wide-angle atomic scattering signal~\cite{assembly_effect}.
Polycrystalline copper aggregates probe both size scaling and computational performance in realistic atomistic microstructures generated from simulations.
Halloysite nanotubes test the method on non-periodic curved geometries with multi-component chemistry.
Colloidal nanoparticle assemblies demonstrate the length-scale independence of the approach by extending its application beyond the atomic scale to simulate small-angle scattering during crystallization.
Across all four cases, AES-Debye yields high-resolution powder profiles, and partial PDFs where relevant, with high numerical fidelity, demonstrating how its computational advantage translates into practical capabilities for materials systems of direct experimental interest.

\subsection{Nanocube superlattices}

To illustrate the ability of the AES-debye to probe hierarchical self-assembly, we computed diffraction profiles for finite nanocube superlattices.
Such assemblies can require specialized treatment for each case, whereas the DSE naturally captures coherence across multiple length scales.

We constructed truncated nanocubes with a 5\,nm edge length ($\approx 20\%$ truncation) and arranged them on a simple cubic lattice with a surface-to-surface separation of 0.5\,nm to mimic a self-assembled superlattice (\cref{fig:assembly_combined}a)
The diffraction profile of the full assembly differs markedly from that of an isolated nanocube (\cref{fig:assembly_combined}b).
Because the particles are small and closely packed, interparticle correlations affect not only the low-$Q$ superlattice reflections but also the wide-angle scattering signal.
The high-frequency serrations in the background are expected finite size features of the idealized nanocubes and are enhanced by the absence of size dispersity.

\cref{fig:assembly_combined}b also shows that increasing the size of the primary nanocrystals reduces the influence of the assembly on the wide angle profile.
Introducing positional and rotational disorder into the superlattice similarly suppresses these assembly-induced features, bringing the profile closer to that of the isolated particle.
Together, these results show that AES-Debye can resolve how hierarchical ordering and disorder shape total scattering signatures in finite nanoparticle assemblies.

\subsection{Polycrystalline copper sample}

To demonstrate the performance of the final implementation on realistic large-scale microstructures, we computed diffraction profiles for a series of polycrystalline Cu models.  
Each structure consisted of multiple randomly oriented grains generated with the constrained modified Voronoi tessellation (CMVT) method~\cite{polygrains} and subsequently relaxed by molecular dynamics (\cref{fig:poly_combined}a) ~\cite{cusample}.  
Time-averaged atomic positions were used as input.
Calculations were performed on 64 CPU nodes of the Fritz cluster (4{,}608 cores total) using the local PDF mode and 25 cells per spatial direction.

\cref{fig:poly_combined}b shows the resulting powder diffraction profiles for samples with different numbers of grains, normalized by atom count.
Clear differences appear in the background and in relative peak intensities, indicating that grains statistics influence the averaged profile.  
A more systematic analysis of the number of grains required to reproduce a representative polycrystalline diffraction pattern is left for future work.

\cref{tab:runtimes} details the corresponding runtimes. As expected, the cost follows approximately quadratic scaling, consistent with the $O(N^2)$ pair distance computation. The largest sample was completed in approximately 52 minutes, with a sustained performance of about 280\,MPd\,s$^{-1}$\,core$^{-1}$ across the benchmark set.

\begin{table}[h]
    \centering
    \begin{tabular}{|l|c|c|c|}
        \hline
        \textbf{\# grains} & 500 & 1000 & 2000 \\
        \hline 
        \textbf{\# atoms (million)} & 22.4 & 44.5 & 90.1 \\
        \hline \hline
        \textbf{Time (s)} & 197 & 766 & 3159 \\
        \hline
    \end{tabular}
    \caption{Corresponding runtimes for varying numbers of grains and atoms.}
    \label{tab:runtimes}
\end{table}

For comparison, we estimate that Rose-X, would require roughly twenty times longer for the largest case even under idealized linear scaling assumptions.
In practice, such scaling is difficult to achieve at high core counts, so the advantage of the present implementation is expected to be larger.

\subsection{Halloysite nanotubes}

Halloysite is a naturally occurring aluminosilicate clay mineral with a tubular microstructure formed by rolling kaolinite layers.  
Halloysite nanotubes (HNTs) typically have lengths of 1--15\,$\mu$m and diameters of 10--100\,nm, which makes them attractive for applications in drug delivery, catalysis, and nanotechnology~\cite{halloymultifarious,fizir2018halloysite,yanghalloy}.  
Their hollow interiors enable encapsulation, while their crystalline framework remains accessible to X-ray diffraction analysis.
Molecular dynamics simulations are also widely used to investigate their mechanical and thermal behavior~\cite{HeidariPebdani2023HalloyMD,prishchenko2018molecularHalloyMD,gianni2023newHalloyMD}.
    
The curved, non-periodic geometry of HNTs is challenging for reciprocal space methods, making the DSE a natural alternative.  
We therefore use AES-Debye to examine how structural parameters, particularly tube diameter, influence the powder diffraction profile.

Atomistic models were generated by rolling kaolinite sheets into tubes of varying diameters, as illustrated in \cref{fig:halloysite_combined}a.  
Larger tubes can contain hundreds of millions of atoms and therefore require an optimized implementations to remain computationally tractable.

\cref{fig:halloysite_combined}b shows the resulting diffraction intensity as a function of $Q$ and tube diameter.
Clear shifts in peak positions and peak shapes demonstrate the sensitivity of the scattering pattern to tube geometry and illustrate how such variations can appear in experimental data.

This example also highlights support for multi-component systems.  
Each halloysite model contains three atomic species (O, Si, and Al), giving rise to six distinct partial PDFs: O--O, O--Si, O--Al, Si--Si, Si--Al, and Al--Al.  
These partial contributions are weighted by the corresponding atomic form factors and summed to obtain the final powder diffraction profile.

\subsection{Colloidal nanoparticles}

Since the Debye scattering equation is formulated strictly in terms of pairwise distances, it is fundamentally independent of the absolute length scale of the system.
This allows the method to be readily adapted for studying self-assembly in colloidal crystals and simulating their corresponding small angle scattering patterns.
To demonstrate this capability, we modeled a macroscopic colloidal assembly comprising 200{,}000 nanoparticles.
Rather than modeling atomistic details, each nanoparticle is represented as a uniform coarse-grained sphere with a diameter of 200\,nm (\cref{fig:colloidal_combined}a).
The ensemble was then compressed within a spherical confinement to induce crystallization, allowing us to track the structural evolution starting from an initial disordered state through subsequent, highly ordered assembly stages.

We computed the scattering profiles for these large scale configurations specifically in the small angle region, where superlattice features typically emerge (\cref{fig:colloidal_combined}b).
By multiplying the resulting interparticle structure factor by the single particle form factor, we obtained the complete scattering profiles representing different stages of the crystallization process.
As the system transitions into a crystalline assembly, distinct Bragg peaks begin to emerge and superimpose over the underlying small angle form factor.
These emerging peaks directly reflect the development of long range periodic order during self-assembly.
Ultimately, this application highlights how AES-Debye can be seamlessly extended beyond the atomic scale, providing a powerful computational tool for investigating mesoscale and macroscopic soft-matter systems.


\section{Conclusions and outlook}

We have presented AES-Debye, a modern high-performance implementation for computing powder diffraction profiles from the Debye scattering equation.
By combining a PDF-based formulation with bin center correction, precision-aware integer accumulation, cache-friendly data structures, and a cell list based domain decomposition, the method improves memory locality and reduces latency, particularly for large and disordered systems compared to other implementations.
A hybrid OpenMP/MPI/CUDA design enables efficient execution on CPUs and GPUs, while distributed memory parallelization sustains performance across nodes with only minimal communication overhead.

The method is numerically robust over a wide range of system sizes and disorder levels.
In particular, bin center correction removes nonphysical negative intensities and improves the fidelity of the diffuse background in the high-$Q$ region.
Because PDFs are constructed internally, the same framework also provides high-resolution PDFs for downstream structural analysis, where fine binning and numerical accuracy are essential.

Applications to nanocubes superlattices, polycrystalline copper, halloysite nanotubes and colloidal nanoparticles further demonstrate that AES-Debye can treat large, compositionally complex systems with irregular geometries, enabling high-resolution total scattering calculations for realistic materials models.
In the benchmarks presented here, AES-Debye achieves state-of-the-art throughput on current hardware and substantially outperforms existing tools, including Rose-X and DebyeCalculator, for comparable problem sizes while preserving numerical accuracy.

Despite these advances, evaluation of the DSE remains inherently $O(N^2)$, and extreme disorder can degrade memory locality.
Several extensions remain of interest, including a HIP/ROCm backend for AMD GPUs, and adaptive or mixed precision schemes with explicit error control.
Optional instrument effects, such as resolution broadening, absorption, diffraction geometry, goniometer error can help incorporate other experimental conditions into the profile.
More broadly, refinement workflows such as PDF-based fitting can also exploit the high-resolution PDFs produced internally.

\section*{Code availability}

The library can be found at: \url{https://gitlab.cs.fau.de/iq23adyz/debye}

The library is released as open source under the GNU General Public License (GPLv3) and provides: 
\begin{itemize}
    \item \textbf{Python interface.} A user-friendly API supports flexible data loading and seamless integration into existing workflows.
    Example scripts are included.
    The bindings are implemented with \texttt{pybind11}~\cite{pybind11}.

    \item \textbf{CLI interface.} A command line tool enables rapid computations and accepts all structure formats supported by the ASE Python library~\cite{ase-paper}.

    \item \textbf{LAMMPS plugin.} A command style plugin for LAMMPS reads atomic positions during simulations and writes the computed profiles to user specified locations.
\end{itemize}

Users may also link the C++ library directly; the C++ API closely mirrors the Python API (see the LAMMPS plugin documentation for details).

The library can be installed via the Python package manager \texttt{pip}.
We employ \texttt{scikit-build-core} with CMake to automatically detect and enable MPI and GPU support when the requisite hardware and libraries are present.
Additional component details and installation instructions are provided on the project’s GitHub page (available on publication).

\section*{Funding Declaration}

This work was supported by the Deutsche Forschungsgemeinschaft (DFG, Grant No.\ LE4543/2-1) and by the Competence Network for Scientific High Performance Computing in Bavaria (KONWIHR, Grant No.\ Z.4-M7635.1/23/15).  
We gratefully acknowledge computational resources provided by the Erlangen National High Performance Computing Center (NHR@FAU) under the NHR project CRC1411D04.

\section*{Author contributions}

N.P. implemented the code, developed optimization strategies, performed testing, curated data, and drafted the manuscript. M.W. developed the CPU-based high-performance computing implementation, and S.K. supported the GPU-based implementation. A.L. conceived the method and contributed the earlier Rose-X software. M.E. and A.L. supervised the project, secured funding, and revised the manuscript. All authors contributed to editing and approved the final version.

\section*{Competing interests}

The authors declare no conflict of interest.

\printbibliography

\renewcommand{\thesection}{\Alph{section}}
\setcounter{section}{0}
\section*{Supporting Information}

\section{Additional optimization strategies evaluated}

We also assessed several alternative optimization strategies that were not adopted in the final implementation because they did not provide a favorable performance--accuracy tradeoff.

\begin{itemize}
    \item \textbf{SIMD vectorization.}
    We explored single-instruction, multiple-data (SIMD) vectorization on CPUs to accelerate bin computations.
    In practice, the SIMD implementation performed worse than the scalar baseline, likely because the kernel is dominated by memory-latency and because there is no native vectorized integer square root for \texttt{int64}; emulation or type conversion eliminated all potential gains.

    \item \textbf{Reduced PDF resolution.}
    We tested whether decreasing the number of bins could improve performance.
    Although this produced modest speedups for crystalline systems, it introduced noticeable errors in the diffraction profiles and did not improve runtimes for disordered systems.
    Because the accuracy loss outweighted the limited benefits, this strategy was not pursued further.

    \item \textbf{Cell lists on the GPU.}
    We also evalulated cell-list-based decomposition on GPUs.
    This did not improve performance: in the absence of an L3 cache and given the already high memory bandwidth of the device, the added indirection and bookkeeping outweighed any gain from improved locality.
    On smaller GPUs, lightweight sorting of atoms by cell index produced only minor benefits and may be useful only in more constrained settings.
\end{itemize}

These results further motivated the emphasis on domain decomposition, cache locality optimizations, and architecture-specific parallelization in the final AES-Debye implementation.
\end{document}